\documentclass{aa}
\usepackage{graphicx}
\usepackage{txfonts}
\usepackage{placeins}
\usepackage[]{hyperref}
\hypersetup{colorlinks=true, urlcolor=blue, citecolor=cyan, pdfborder={0 0 0}}

\begin{document}

\title{An analysis of the most distant catalogued open clusters}
\subtitle{Re-assessing fundamental parameters with Gaia EDR3 and
\texttt{ASteCA}\thanks{
Table~\ref{tab:results} is only available in electronic form
at the CDS via anonymous ftp to cdsarc.u-strasbg.fr (\url{130.79.128.5})
or via \url{http://cdsweb.u-strasbg.fr/cgi-bin/qcat?J/A+A/}}}

\author{G. I. Perren\inst{1,3}
      \and
      M. S. Pera\inst{1,3}
      \and
      H. D. Navone\inst{2,3}
      \and
      R. A. Vázquez\inst{1,4}
      % \fnmsep\thanks{Just to show the usage
      % of the elements in the author field}
}

\institute{Instituto de Astrof\'isica de La Plata (IALP-CONICET), La Plata,
Argentina\\
\email{gabrielperren@gmail.com}
\and
Instituto de Física de Rosario (CONICET-UNR), 2000 Rosario, Argentina
\and
Facultad de Ciencias Exactas, Ingeniería y Agrimensura (UNR), 2000 Rosario,
Argentina
\and
Facultad de Ciencias Astronómicas y Geofísicas (UNLP-IALP-CONICET), 1900 La
Plata, Argentina
% \thanks{The university of heaven temporarily does not
%         accept e-mails}
}
\date{Received September 15, 2021; accepted December 16, 2021}

% \abstract{}{}{}{}{} 
% 5 {} token are mandatory
 
\abstract
% context heading (optional)
% {} leave it empty if necessary  
{Several studies have been presented in the last few years applying some kind of
automatic processing of data to estimate the fundamental parameters of open
clusters. These parameters are later on employed in larger scale analyses, for
example the structure of the Galaxy's spiral arms.
The distance is one of the more straightforward parameters to estimate, yet
enormous differences can still be found among published data. This is
particularly true for open clusters located more than a few kpc away.}
% aims heading (mandatory)
{
We cross-matched several published catalogues and selected the twenty-five most
distant open clusters ($>$9000 pc). We then performed a detailed analysis of
their fundamental parameters, with emphasis on their distances, to determine the
agreement between catalogues and our estimates.}
% methods heading (mandatory)
{Photometric and astrometric data from the Gaia EDR3 survey was employed. The
data was processed with our own membership analysis code (pyUPMASK), and our
package for automatic fundamental cluster's parameters estimation
(\texttt{ASteCA}).}
% results heading (mandatory)
{We find differences in the estimated distances of up to several kpc
between our results and those catalogued, even for the catalogues that show the
best matches with \texttt{ASteCA} values. Large differences are also found for
the age estimates. As a by-product of the analysis we find that
vd Bergh-Hagen 176 could be the open cluster with the largest heliocentric
distance catalogued to date.}
% conclusions heading (optional), leave it empty if necessary
{Caution is thus strongly recommended when using catalogued parameters of open
clusters to infer large-scale properties of the Galaxy, particularly for those
located more than a few kpc away.}

\keywords{
  Methods: statistical --
  Galaxies: star clusters: general --
  (Galaxy:) open clusters and associations: general --
  Techniques: photometric --
  Astronomical databases: miscellaneous
}

\maketitle

% =============================================================================
\section{Introduction}

 The unprecedented amount of high precision data for parallaxes, proper motions,
 and photometry provided by the Gaia mission in successive
 deliveries~\citep[DR2 and EDR3,][]{Gaia_2016,Gaia_EDR3} offers us a unique
 opportunity to estimate the fundamental parameters of open clusters (OC): metal
 content, age, total mass, binary fraction, distance,  and extinction.
 The arrival of new techniques for analyzing massive data combined with the
 increasing data precision promise more reliable results than those obtained
 with the old techniques. The latter were mostly based on the visual inspection
 of their color-magnitude diagrams and isochrone fittings \citep{Phelps1994}, or
 on direct comparison with HR diagrams of synthetic clusters \citep{Siess1997}.
 Automated processes such as the one applied by \cite{Kharchenko_2012} have also
 played an important role in determining cluster parameters in a massive way.
 The continuous increase of high quality data is a defying circumstance
 where a variety of analysis are being considered including artificial
 neural networks~\citep{Cantat_2020}, combined with new strategies for
 determining cluster memberships \citep{Krone2014,Cantat-Gaudin_2018} or
 dynamical
 evolution analysis as the one applied by \citep{Gregorio_2015}.

 The intrinsic value of studying OCs has been profusely described in several
 opportunities, we  give here only a brief enumeration of the importance
 of these objects. The oldest OCs
 allow us to investigate the height and radial extension of the galactic disk,
 old OCs tell us about the chemical history (age-metallicity relation), the
 mixing processes (radial metallicity gradient), and the processes of cluster
 destruction by interaction with other populations of the
 galaxy~\citep{Friel1995,Tosi_2004,Lamers_2005}.
 The youngest OCs, on the other hand, are not only used as laboratories to
 investigate stellar evolution~\citep[they allow studying in detail the boundary
 conditions necessary to create new generations of stars, ][]{Lada2003} but are
 also routinely employed in the analysis of the Milky Way's
 structure~\citep{Loktin_1992,Moitinho_2006,Vazquez2008,Moitinho_2010}
 becoming particularly useful in the tracing of spiral
 arms~\citep{carraro_2013,Molina_2018}.
 Young OCs are arranged along the galactic disk, where the strong visual
 absorption and the contamination by field stars very often prevent observing
 stars in the lower part of their main sequence.
 The situation is not much better for the older OCs which do not
 have very luminous stars in the main sequence, although they do in the giant
 branch. Stars in the lower part of the main sequences, as well as those
 belonging to the giant branch, share similar photometric characteristics with
 field stars making it rather difficult to unravel to which population each star
 belongs~\citep{Hayes_2015}.
 The situation worsens as the distance to the older OCs increases because the
 limiting magnitude increases, which results in only a small portion of
 the lower part being visible. But it is not only the photometric
 data dimensions that are disturbed by distance. The proper
 motions of distant OCs are
 extremely difficult to separate from those characterizing the field
 population against which we see them projected, therefore introducing an
 additional degree of confusion in determining memberships.

 Our interest in this current article is twofold. On the one hand, it is focused
 on reexamining the distances and properties of the most distant OCs cataloged
 so far in our Galaxy. A total of 25 clusters that satisfy this requirement were
 found inspecting four different recognized catalogs/databases, as we will
 explain below.
 However, these catalogs display enormous differences in the estimated distances
 and ages. In part, these differences for the same catalogued object may be due
 to the varying techniques used to perform the analysis, combined with
 the problem of the very large distance at which they are located.
 %
 % The ages for most of these OCs, for example, are presumed to be larger than one
 % billion years. But in other catalogs these
 % same clusters have relatively small ages and shorter distances as well.
 We want to contribute to the task of resolving these differences.
 On the other hand, we want to test our new membership estimation technique
 pyUPMASK\footnote{\url{https://github.com/msolpera/pyUPMASK}}~\citep{Pera_2021}
 in combination with
 \texttt{ASteCA}\footnote{\url{http://asteca.github.io/}}~\citep{Perren_2015},
 on clusters with proper motions that are not easily distinguishable from that
 of surrounding stars, composed of a small number of members, and with
 non-trivial sequences in the photometric space.\\

 This article is structured as follows. In Sect.~\ref{sec:cat_clust_data} we
 introduce the stellar cluster catalogues, the clusters selected to be
 analyzed (crossed-matched from those catalogues), and the photometric and
 astrometric data used to perform the analysis.
 Sect.~\ref{sec:clust_analy} presents the methods employed in the study of all the
 clusters. The comparison of the estimated parameters with the catalogued
 values for each cluster is done in Sect.~\ref{sec:results}. Finally,
 conclusions are highlighted in Sect.~\ref{sec:conclusions}.

% =============================================================================
\section{Catalogues, clusters, and data}
 \label{sec:cat_clust_data}

 We selected four catalogues to cross-match and subsequently use to identify the
 most distant clusters: \citet[][New Catalog of Optically Visible Open Clusters
 and Candidates, hereinafter OC02]{Dias_2002},~\citet[][hereinafter
 WEBDA\footnote{\url{https://webda.physics.muni.cz/}}]{Netopil_2012},
 \citet[][Milky Way Star Clusters Catalog, hereinafter MWSC]{Kharchenko_2012},
 and~\citet[][hereinafter CG20]{Cantat_2020}.
 The first two (OC02 and WEBDA) are compilations of open clusters' fundamental
 parameters from the literature. They contain around 1700 (WEBDA) and 2100 
 (OC02) entries, and are heavily used in the field of open cluster research.
 The parameter values in both catalogues are heterogeneous, being compiled from
 various sources.
 The MWSC catalog is the largest one ($\sim$3000 entries) and, similarly to the
 CG20 catalog ($\sim$2000 entries), is composed of homogeneous fundamental
 parameter values obtained for all its entries.
 The method employed by the authors of the MWSC catalog is a semi-automated
 isochrone fit applied on clusters and candidate clusters, while the
 CG20 catalog was generated employing an artificial neural network on
 verified clusters only (trained on parameter values taken from the
 literature).

 Since we are interested in the open clusters most distant from the Sun, we
 select from these cross-matched catalogues those that are located at a
 distance of 9000 pc or more in either of them. This is an arbitrary value that
 results in enough clusters to draw general conclusions, but not too many that
 would impede their detailed analysis. The final twenty-five clusters that were
 studied in this work are shown in Table~\ref{tab:clusters}.\\

 \begin{table*}
 \caption{Selected open clusters with a catalogued distance $\geq$9000 pc,
 ordered by right ascension. The ages are expressed as the logarithm, and the
 distances are in parsec. In parenthesis, the short names used for the clusters
 throughout the article in tables and figures. Clusters with no distances below
 the 9000 pc limit in any of the catalogs are marked with boldface.}
 \label{tab:clusters}
 \centering
 % \begin{tabular}{lccccccccc}
 \begin{tabular}{llrrrrrrrrr}
 \hline\hline
 Cluster & $\alpha_{2000}$  & $\delta_{2000}$ & \multicolumn{2}{c}{OC02} &
 \multicolumn{2}{c}{CG20} & \multicolumn{2}{c}{WEBDA} & \multicolumn{2}{c}
 {MWSC}\\
   &  &  & age & dist & age & dist & age & dist & age & dist \\
 \hline
  Berkeley 73 (BER73)             & 95.5   & -6.35  & 9.18  & 9800  & 9.15 & 6158  & 9.36  & 6850  & 9.15  & 7881 \\
  Berkeley 25 (BER25)             & 100.25 & -16.52 & 9.70  & 11400 & 9.39 & 6780  & 9.60  & 11300 & 9.70  & 11400 \\
  Berkeley 75 (BER75)             & 102.25 & -24.00 & 9.60  & 9100  & 9.23 & 8304  & 9.48  & 9800  & 9.30  & 6273 \\
  Berkeley 26 (BER26)             & 102.58 & +5.75  & 9.60  & 12589 & -    & -     & 9.60  & 4300  & 8.71  & 2724 \\
  Berkeley 29 (\textbf{BER29})    & 103.27 & 16.93  & 9.03  & 14871 & 9.49 & 12604 & 9.03  & 14871 & 9.10  & 10797 \\
  Tombaugh 2 (TOMB2)              & 105.77 & -20.82 & 9.01  & 6080  & 9.21 & 9316  & 9.01  & 13260 & 9.01  & 6565 \\
  Berkeley 76 (BER76)             & 106.67 & -11.73 & 9.18  & 12600 & 9.22 & 4746  & 9.18  & 12600 & 8.87  & 2360 \\
  FSR 1212 (F1212)                & 106.94 & -14.15 & -     & -     & 9.14 & 9682  & -     & -     & 8.65  & 1780 \\
  Saurer 1 (\textbf{SAU1})        & 110.23 & +1.81  & 9.70  & 13200 & -    & -     & 9.85  & 13200 & 9.60  & 13719 \\
  Czernik 30 (CZER30)             & 112.83 & -9.97  & 9.40  & 9120  & 9.46 & 6647  & 9.40  & 6200  & 9.20  & 6812 \\
  Arp-Madore 2 (\textbf{ARPM2})   & 114.69 & -33.84 & 9.34  & 13341 & 9.48 & 11751 & 9.34  & 13341 & 9.34  & 13338 \\
  vd Bergh-Hagen 4 (\textbf{BH4}) & 114.43 & -36.07 & -     & -     & -    & -     & 8.30  & 19300 & -     & - \\
  FSR 1419 (F1419)                & 124.71 & -47.79 & -     & -     & 9.21 & 11165 & -     & -     & 8.38  & 7746 \\
  vd Bergh-Hagen 37 (BH37)        & 128.95 & -43.62 & 8.84  & 11220 & 8.24 & 4038  & 8.85  & 2500  & 7.50  & 5202 \\
  ESO 092 05 (E9205)              & 150.81 & -64.75 & 9.30  & 5168  & 9.65 & 12444 & 9.78  & 10900 & 9.30  & 5168 \\
  ESO 092 18 (E9218)              & 153.74 & -64.61 & 9.02  & 10607 & 9.46 & 9910  & 9.02  & 607   & 9.15  & 9548 \\
  Saurer 3 (SAU3)                 & 160.35 & -55.31 & 9.30  & 9550  & -    & -     & 9.45  & 8830  & 9.30  & 7075 \\
  Kronberger 39 (KRON39)          & 163.56 & -61.74 & -     & 11100 & -    & -     & -     & -     & 6.00  & 4372 \\
  ESO 093 08 (E9308)              & 169.92 & -65.22 & 9.74  & 14000 & -    & -     & 9.65  & 3700  & 9.80  & 13797 \\
  vd Bergh-Hagen 144 (BH144)      & 198.78 & -65.92 & 8.90  & 12000 & 9.17 & 9649  & 8.90  & 12000 & 9.00  & 7241 \\
  vd Bergh-Hagen 176 (BH176)      & 234.85 & -50.05 & -     & -     & -    & -     & -     & 13400 & 9.80  & 18887 \\
  Kronberger 31 (\textbf{KRON31}) & 295.05 & +26.26 & -     & 11900 & -    & -     & -     & -     & 8.50  & 12617 \\
  Saurer 6 (SAU6)                 & 297.76 & +32.24 & 9.29  & 9330  & -    & -     & 9.29  & 9330  & 9.20  & 7329 \\
  Berkeley 56 (\textbf{BER56})    & 319.43 & +41.83 & 9.60  & 12100 & 9.47 & 9516  & 9.60  & 12100 & 9.40  & 13180 \\
  Berkeley 102 (BER102)           & 354.66 & +56.64 & 9.50  & 9638  & 9.59 & 10519 & 8.78  & 2600  & 9.14  & 4900 \\
 \hline
 \end{tabular}
 \end{table*}

 Our full list initially consisted of thirty-eight open clusters; eleven of
 these were found only in the MWSC catalog with distances larger than 9000 pc.
 These are either listed with substantially smaller distances in the other
 catalogs, or too sparse and/or dubious. Hence these clusters were removed from
 the cross-matched list.
 Two other clusters were also removed from the initial list : Shorlin 1 
 ($\alpha_{2000}$=166.44, $\delta_{2000}$=-61.23) and FSR0338
 ($\alpha_{2000}$=327.93, $\delta_{2000}$=55.33). The latter appears in WEBDA and
 MWSC at a distance of 12600 pc and 5600 pc respectively, while the former is
 listed only in MWSC with a distance of 14655 pc. Shorlin1 is studied
 in~\cite{Carraro_2009} and~\cite{Turner_2012}; in both cases the authors
 conclude that this is not a real cluster but a grouping of young stars.
 % Shorlin 1
 %
 % Carraro & Costa (2009)
 % > We also present evidence that this extremely distant group, formerly assumed
 % to be a star cluster (Shorlin 1), is a diffuse, young population, typically
 % found in spiral galaxies.
 %
 % Turner (2012)
 % > ...star counts also imply that Shorlin 1 is unlikely to represent a true
 % cluster, but is instead merely the trapezium-like remains of a previously-
 % bound cluster with an implied age of only a few Myr, according to the likely
 % presence of O-type stars within its boundaries.
 %
 FSR0338 is analyzed in~\cite{Froebrich_2010} where a distance of 6000 pc is
 assigned, but with large uncertainties.
 % Froebrich et al. (2010)
 % > Th high probability members of this newly identified cluster seem to indicate
 % an object with an age of 0.4 Gyrs at a distance of 6 kpc. The scatter in the
 % data points suggests larger than normal uncertainties for the parameters.
 %
 In both cases we find no evidence of a true stellar cluster in these regions.
 We base our conclusion on two findings. First, the large proper motions
 dispersion of the stars that occupy the overdensity around the central
 coordinates assigned to either object. Second, the lack of a clear sequence in
 their respective color-magnitude diagrams (CMD hereinafter). These two clusters
 are thus also discarded from further analysis.

 Most of the twenty-five selected clusters are located in the Third Quadrant
 with all of them in the latitude range of $\left[-12^{\circ}, 8^{\circ}\right]$,
 relatively close to the galactic plane. The final list thus contains 24
 clusters present in the MWSC catalog, 21 in WEBDA, 19 in OC02, and 16 in
 CG20.\\

 There are two other major works where a large catalog of analyzed open clusters
 is presented: \cite{Lui_2019} and \cite{Dias_2021}. The former does not contain
 clusters with such large distances, and was not used. The latter lists
 only four clusters that are also present in our set of twenty-five selected
 clusters. None of their distances comply with our selection filter, hence this
 database was not included. This fact notwithstanding, their distance values will
 be mentioned in the discussion of the results in Sect.~\ref{sec:results}.\\

\begin{figure*}
  \resizebox{\hsize}{!}{\includegraphics[]{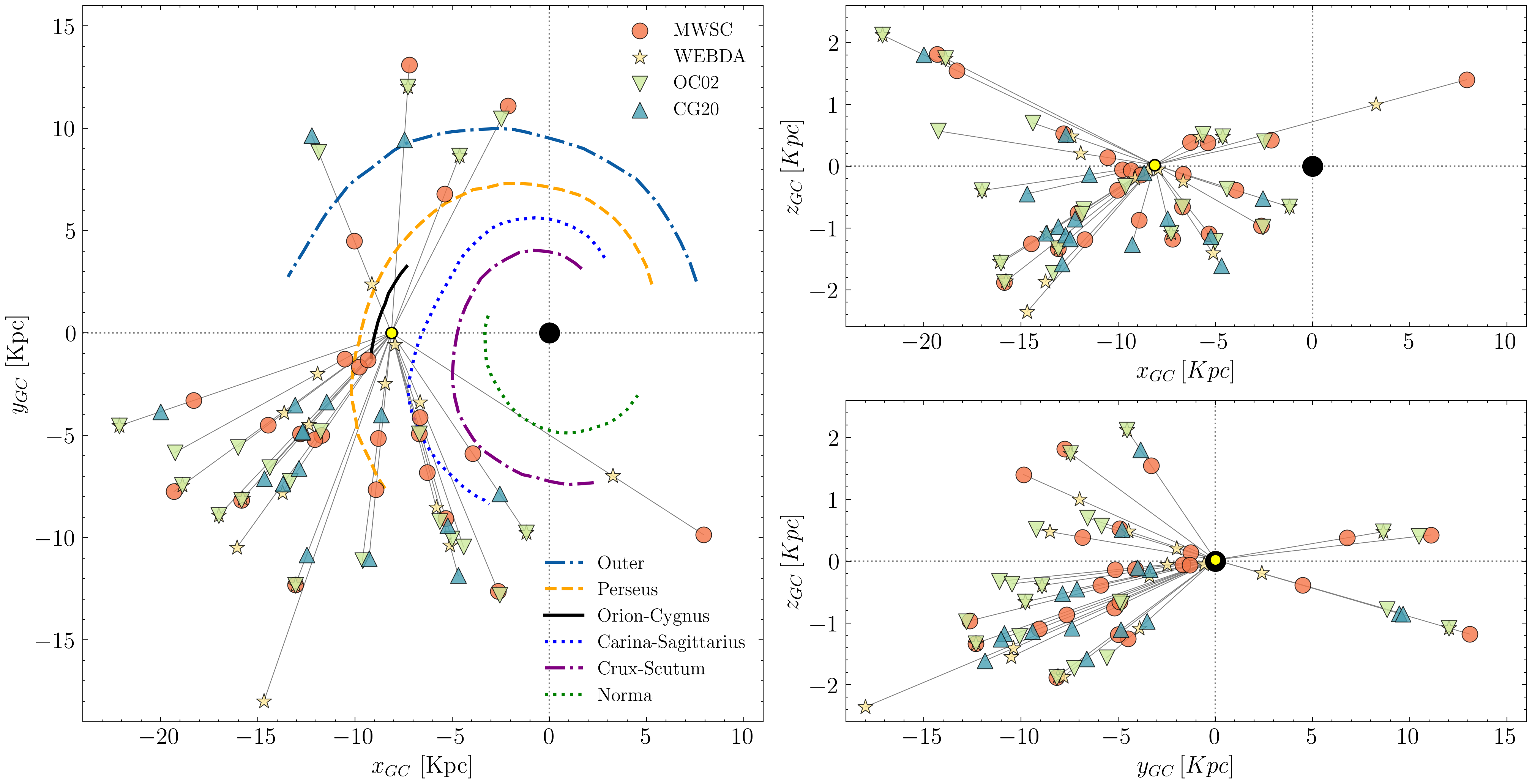}}
  \caption{Left: position of the twenty-five clusters selected from the four
    catalogs mentioned in the text, on a face-on view of the Milky Way. The Sun
    and the center of the Galaxy are marked with a yellow filled circle and a
    black filled circle, respectively. Right, top and bottom: edge-on views,
    same color and marker conventions as above.
    Sight lines shown in grey for each cluster.}
  \label{fig:MWmap}
 \end{figure*}

 Data from Gaia EDR3~\citep{Gaia_2016,Gaia_EDR3} was retrieved for a box of 20
 arcmin of length around the central coordinates for all the clusters. We
 employed equatorial coordinates, parallax, proper motions, and photometry
 ($G$, $G_{BP}-G_{RP}$) from these survey.
 In Fig~\ref{fig:MWmap} we show the twenty-five selected clusters for each of
 the four catalogues, positioned on the face-on view of the Galaxy (top), and
 two edge on views (center, bottom). The spiral arms are those presented
 in~\cite{Momany_2006}. The large dispersion for the distances
 assigned to each cluster in different catalogs is clearly visible, where
 ideally the position of all the clusters would overlap for the four catalogs.\\

 In what follows we will only show the figures for a single representative
 cluster (Berkeley 29) to avoid the clutter and improve the readability of the
 article. The plots for the remaining clusters can be found in the Appendix.

% =============================================================================
\section{Cluster analysis}
 \label{sec:clust_analy}

 \subsection{Structural analysis}

  The first step in the cluster analysis is the estimation of their structural
  properties, i.e. center coordinates and limiting radius. Although centers and
  diameters are present in (some of) the catalogues, not all of these values are
  correct. We use our \texttt{ASteCA} package throughout this
  work to perform the structural and fundamental parameters analysis. We have
  applied this tool to the study of hundreds of clusters in previous articles,
  with excellent results~\citep{Perren_2017,Perren_2020}.

  The center values are obtained applying a two-dimensional kernel density
  analysis (KDE) on each of the cluster's coordinates. This method assigns the
  center of the cluster to the point with the largest density in the frame. As
  shown in previous articles \citep{Perren_2015,Perren_2017,Perren_2020}, this
  approach is robust even when applied on frames with star densities that
  are very much non-uniform. See for example the case of van den Bergh-Hagen 37,
  shown in Fig.~\ref{fig:12struct}.

  A King's profile~\citep{King_1962} fit is performed on the radial
  density profile (RDP hereinafter) of each cluster to estimate their core and
  tidal radii ($r_{c}$, $r_{t}$). The adopted radius $r_{a}$ is the limiting
  distance from the center used to define the studied cluster region for each
  cluster. These radii are estimated applying a process that compares the
  ratio of the approximated number of true members for increasing radii values,
  with the number of stars in a concentric ring centered on each radius. The
  approximated number of members is obtained as the total number of stars within
  the radius, minus the expected number (field density times circle area). This
  method produces an overdensity around the value where the radial density
  approaches the field density, maximizing the contrast between members included
  within the radius and contaminating field stars. The method is also useful
  for heavily contaminated cluster, and/or clusters with very few true
  members. All radii values are shown in Table~\ref{tab:radii}.\\

  The adopted radius $r_{a}$ is on average 50\% smaller than the
  tidal radius (see Table~\ref{tab:radii}). This allows us to alleviate the
  issue of field star contamination, while ensuring that only a small number of
  true members (cluster stars located as far from the center as the tidal
  radius) are lost.
  The fraction of lost members can be estimated integrating King's profile. This
  fraction depends on the concentration of the cluster ($r_{t}/r_{c}$) and the
  value of the adopted radius as a fraction of the tidal radius ($r_{a}/r_{t}$).
  In our case, less than 20\% of the members could be lost in a worst case
  scenario. Since these are clusters that are strongly contaminated 
  (particularly in the parallax and proper motions spaces), the trade off
  between losing a small portion of members and improving the contrast of the
  true members over the field noise, is positive.
  The $r_{a}$ values used in our analysis being smaller than the tidal radius,
  means that the total estimated mass for each cluster shown in
  Sect.~\ref{sec:results} must be thought of as a lower limit.\\

\begin{figure*}
   \resizebox{\hsize}{!}{\includegraphics[]{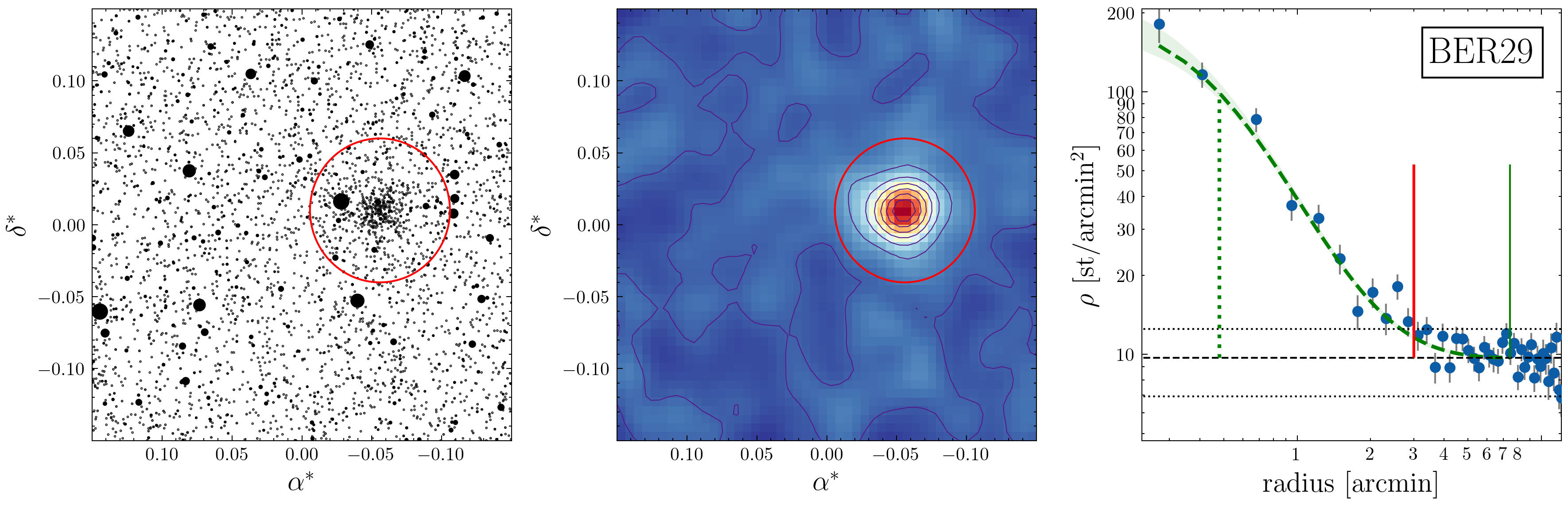}}
   \caption{Left: analyzed $20^{\prime} \times 20^{\prime}$ arcmin frame with the
   estimated cluster region enclosed in a red circle. Center: same frame but
   shown as as two-dimensional density map. Right: radial density plot in
   logarithmic axis. Details in the body of the article.}
   \label{fig:BER29_struct}
  \end{figure*}

  In Fig.~\ref{fig:BER29_struct} we show the structural analysis, center and
  radii estimation processes, for the cluster Berkeley 29. The asterisks in the
  equatorial coordinates of the left plot indicate that these were shifted
  and transformed so that the center of the frame is located at (0, 0) and to
  remove projection artifacts. The right plot shows the radial density analysis
  where the dashed green line and the shaded green area are the King profile fit
  and its 16th-84th uncertainty region, respectively. The green dotted vertical
  line, solid red vertical line, and solid green vertical line, are the core 
  ($r_{c}$), adopted ($r_{a}$), and tidal ($r_{t}$) radii, respectively. The
  dashed and dotted horizontal black lines are the field density estimate and
  its $\pm1\sigma$ region, respectively. The plots for the remaining cluster can
  be seen in Appendix~\ref{app:struct_analysis}.

 \subsection{Membership and fundamental parameters}
  \label{ssec:fund_pars}

  Before we can estimate the fundamental parameters with \texttt{ASteCA}, we
  need to select the set of most probable members for each cluster. For this
  task we employed our recently developed pyUPMASK algorithm which demonstrated
  a great performance even for very contaminated clusters, outperforming
  UPMASK~\citep{Krone2014} as shown in~\cite{Pera_2021}. Internal tests
  showed that pyUPMASK also out-performs \texttt{ASteCA}'s own membership
  algorithm, hence the reason why we select the former over the latter.

  pyUPMASk requires an input data set composed of $(\alpha, \delta)$ coordinates
  and at least two dimensions of data of any type to estimate the membership
  probabilities. We chose to make use of the proper motion data dimensions only,
  thus excluding photometric and parallax data.
  We made the decision of leaving out these extra data dimensions because,
  although they can be sometimes useful in the process of singling out the most
  probable members, for these type of very distant clusters they tend to add
  more noise than information.
  This is particularly true for the parallax data which
  rapidly tends to zero for stars beyond $\sim$2 kpc, where the parallax values
  for the cluster members become almost indistinguishable from the contaminating
  field stars. The selected clustering method in pyUPMASK was a Gaussian Mixture
  Model, which demonstrated to have the best performance
  in~\citet[][see Sect. 4]{Pera_2021}.

  Once pyUPMASK has assigned membership probabilities to all the stars in the
  frame, we must select the set of stars that most likely belong to the
  cluster (i.e., true members). This selection is performed within the
  cluster region, defined as $r\leq r_{a}$, where $r$ is the distance to the
  cluster's center.
  This step is usually handled by selecting
  an arbitrary cut-off probability value; in~\cite{Cantat_2020} for example
  the authors fix this value to P=70\%. Instead of setting an ad-hoc value, we
  performed an analysis that combines the membership probabilities with the
  stellar density inside and outside of the cluster region. This allows us to
  estimate the number of cluster members expected within the cluster region.
  Combining this number with the membership probabilities given by pyUPMASK we
  select those stars with the largest probabilities within the cluster region,
  such that the resulting total number of members is as close as possible as the
  expected one (i.e., the one obtained through the stellar density analysis).
  Using a physically reasonable number of members not only reduces the
  probability of excluding true members (by only selecting those with the
  largest membership probabilities), it also ensures that the estimation of the
  total mass parameter is properly performed by \texttt{ASteCA}.\\

  After selecting the set of true members for all the clusters as described
  above, we feed this data directly to the final section of our 
  \texttt{ASteCA} package bypassing its internal membership algorithm. 
  The goal of this section is to estimate
  the fundamental parameters: metallicity, age, total mass, fraction of binary
  systems, distance, and extinction. The code uses the ptemcee parallel
  tempering Bayesian inference algorithm~\citep{ptemcee} to sample the
  fundamental parameters' distributions. The likelihood function employed to
  asses the fit between the observed cluster and the synthetic clusters is the
  Bayesian Poisson ratio defined in~\cite{Tremmel_2013}.
  The theoretical isochrones used to generate the
  synthetic clusters used to match the observed clusters are the PARSEC
  tracks~\citep{Bressan_2012}. Priors are uniform for all the parameters using
  the following limiting ranges:

  \begin{itemize}
   \item metallicity ([Fe/H]): [-0.60, 0.30]
   \item logarithmic age: [8, 10.1]
   \item total mass: [1e2, 2e5] $M_{\odot}$
   \item binarity fraction: [0, 1]
   \item distance modulus: [10, 20] mag
   \item $E_{BV}$ extinction: [0, $E_{BV}^{max}$]
  \end{itemize}

  \noindent The maximum value for the extinction priors, $E_{BV}^{max}$, was set
  on a per-cluster basis selecting the values given by the~\cite{Schlegel_1998}
  extinction maps with the re-calibration by~\cite{Schlafly_2011}.
  The logarithmic abundances [Fe/H] were obtained using the approximation given
  in the CMD service for [M/H],\footnote{Assuming [M/H]$\sim$[Fe/H], [Fe/H]$=\log(Z/X)-\log(Z/X)_{o}$, where:
  $(Z/X)_{o}=0.0207; Y=0.2485+1.78Z$.\\
  CMD service: \url{http://stev.oapd.inaf.it/cgi-bin/cmd}} given that the
  PARSEC isochrones are generated using Z.

  \begin{figure*}
   \resizebox{\hsize}{!}{\includegraphics[]{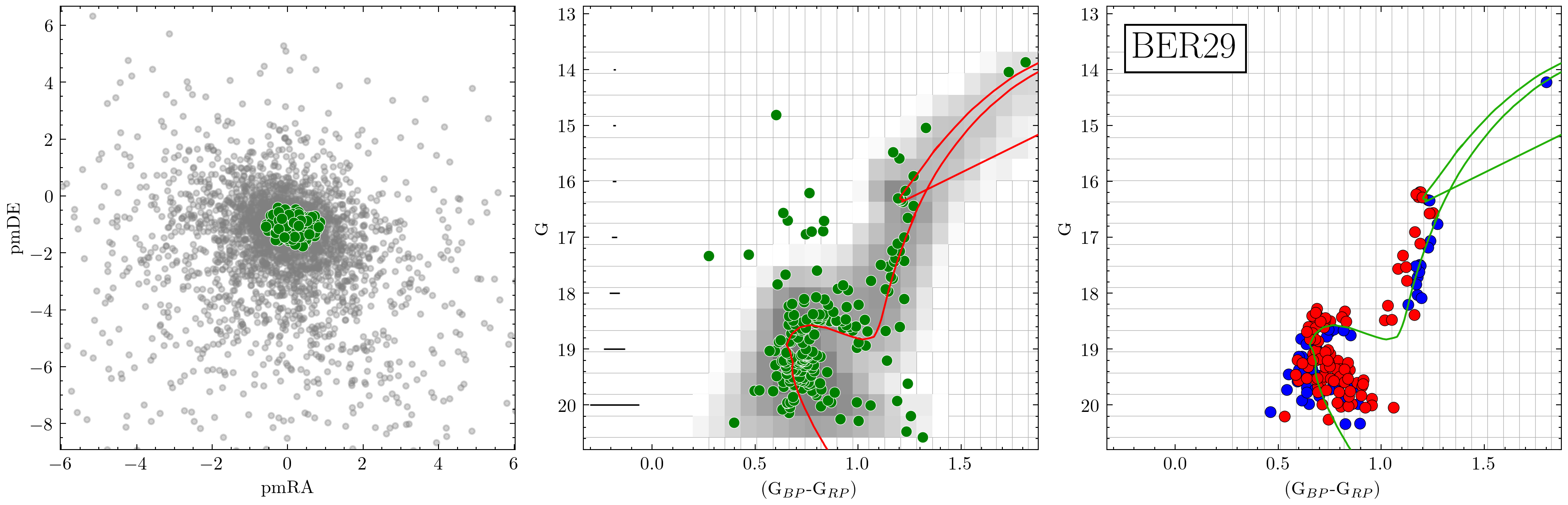}}
   \caption{Left: VPD for stars in the analyzed Berkeley 29
    frame; green and grey circles show the selected true members and the field
    stars, respectively.
    Center: CMD for the cluster's members with the isochrone associated to the
    best synthetic cluster fit drawn in red to guide the eye.
    Right: best synthetic cluster fit found by \texttt{ASteCA} with the same
    isochrone now show in green. Blue and red circles are single and binary
    systems, respectively.}
   \label{fig:BER29_fpars}
  \end{figure*}

  In Fig.~\ref{fig:BER29_fpars} we show the result of the membership
  probabilities estimation done with pyUPMASK, plus the fundamental parameters
  estimation performed by \texttt{ASteCA}. We only show here the plots for
  the cluster Berkeley 29, the remaining clusters can be seen in
  Appendix~\ref{app:fundam_params}.
  The plot on the left shows the vector point diagram (VPD) with the proper
  motion distributions for both the selected clusters members, and the field
  stars. The members are clearly very much embedded within the field stars
  distribution, which is expected for distant clusters. The center plot shows
  the CMD traced by the selected members, and the right plot a sampling of the
  best fit synthetic cluster. The grid in the center and right plots is
  the 2-dimensional binning used to estimate the likelihood, obtained using
  Knuth's rule~\citep{Knuth_2006}. The grey region represents the uncertainty
  in the fit. The isochrone drawn in the center and right plots is associated
  with the synthetic cluster but it is there merely to guide the eye; the fit is
  performed for the CMD of the observed cluster versus the CMD of synthetic
  clusters, not versus theoretical isochrones~\citep[this is further
  explained in:][]{Perren_2015,Perren_2017,Perren_2020}.\\

% =============================================================================
\section{Results and discussion}
 \label{sec:results}

 We present the general results for the
 fundamental parameters contrasted with values taken from the
 aforementioned databases, with particular emphasis on the distances.
 %
 % Finally, in Sect.~\ref{ssec:met_gradient} we analyze our results in the context
 % of the radial metallicity gradient and the age vs metallicity relation.\\
 %
 In Appendix~\ref{app:indiv_clusters} we discuss each cluster
 individually, commenting on the most relevant studies published in the
 literature and how these compare to the results obtained in this article.\\

 Henceforth we employ the default values for the Galactocentric coordinate
 frame given by the astropy
 package\footnote{\url{https://docs.astropy.org/en/stable/coordinates/galactocentric.html}}:

 \begin{itemize}
  \item ICRS coordinates of the Galactic center: (266.4051$^{\circ}$,\\
  -28.936175$^{\circ}$)
  \item Distance from the sun to the Galactic center: 8.122 pc
  \item Distance from the sun to the Galactic midplane: 20.8 pc
  \item Velocity of the sun in the Galactocentric frame as Cartesian velocity
  components: (12.9, 245.6, 7.78) km/s
 \end{itemize}

  \begin{table*}
  \caption{Fundamental parameters estimated with \texttt{ASteCA} for the
  twenty-five analyzed clusters. Sub and supra indexes
  indicate the 16th and 84th percentiles, respectively. The last column
  indicates the number of true members used in the analysis.}
  \label{tab:results}
  \centering
  \begin{tabular}{lccccccccc}
  \hline\hline
  Cluster  & [Fe/H] & $\log{age}$ & $E_{BV}$ & dm$_{\odot}$ & Dist $[kpc]$ & M (M$_{\odot}$) & b$_{fr}$ & N\\
  \hline %\\[.001cm]
    BER73 & $-0.41_{-0.50}^{-0.15}$ & $9.60_{9.27}^{10.02}$ &
    $0.16_{0.06}^{0.35}$ & $13.70_{13.08}^{14.30}$ & $5.49_{4.14}^{7.25}$ &
    $3.3E+03_{2.1E+03}^{5.9E+03}$ & $0.53_{0.22}^{0.81}$ & 103 \\[.2cm]
    BER25 & $-0.20_{-0.49}^{0.00}$ & $9.72_{9.62}^{9.79}$ &
    $0.39_{0.35}^{0.46}$ & $14.34_{14.25}^{14.45}$ & $7.37_{7.08}^{7.76}$ &
    $1.5E+04_{9.1E+03}^{2.2E+04}$ & $0.82_{0.60}^{0.94}$ & 213 \\[.2cm]
    BER75 & $-0.39_{-0.55}^{-0.03}$ & $9.74_{9.68}^{9.84}$ &
    $0.11_{0.05}^{0.16}$ & $14.52_{14.25}^{14.72}$ & $8.03_{7.08}^{8.80}$ &
    $6.8E+03_{3.0E+03}^{1.3E+04}$ & $0.77_{0.19}^{0.96}$ & 95 \\[.2cm]
    BER26 & $0.07_{-0.20}^{0.24}$ & $9.94_{9.79}^{10.06}$ &
    $0.55_{0.50}^{0.59}$ & $13.30_{12.96}^{13.67}$ & $4.57_{3.91}^{5.43}$ &
    $4.6E+03_{2.3E+03}^{8.5E+03}$ & $0.78_{0.45}^{0.95}$ & 76 \\[.2cm]
    BER29 & $-0.21_{-0.32}^{-0.09}$ & $9.57_{9.52}^{9.61}$ &
    $0.07_{0.04}^{0.10}$ & $15.79_{15.67}^{15.88}$ & $14.41_{13.64}^{15.03}$ &
    $1.1E+04_{7.4E+03}^{1.8E+04}$ & $0.56_{0.34}^{0.82}$ & 202 \\[.2cm]
    TOMB2 & $-0.48_{-0.49}^{-0.47}$ & $9.33_{9.31}^{9.34}$ &
    $0.39_{0.39}^{0.40}$ & $14.70_{14.67}^{14.74}$ & $8.73_{8.59}^{8.88}$ &
    $2.1E+04_{1.8E+04}^{2.2E+04}$ & $0.45_{0.40}^{0.49}$ & 845 \\[.2cm]
    BER76 & $-0.11_{-0.34}^{0.02}$ & $9.26_{9.20}^{9.32}$ &
    $0.60_{0.55}^{0.66}$ & $13.66_{13.53}^{13.81}$ & $5.40_{5.08}^{5.77}$ &
    $4.3E+03_{3.0E+03}^{6.8E+03}$ & $0.61_{0.42}^{0.81}$ & 156 \\[.2cm]
    F1212 & $-0.12_{-0.34}^{0.11}$ & $9.10_{9.05}^{9.16}$ &
    $0.66_{0.60}^{0.72}$ & $15.01_{14.85}^{15.20}$ & $10.05_{9.34}^{10.97}$ &
    $5.0E+03_{3.6E+03}^{8.4E+03}$ & $0.51_{0.33}^{0.76}$ & 99 \\[.2cm]
    SAU1 & $-0.08_{-0.34}^{0.16}$ & $9.82_{9.71}^{9.92}$ &
    $0.13_{0.07}^{0.17}$ & $15.46_{15.22}^{15.66}$ & $12.37_{11.07}^{13.54}$ &
    $1.0E+04_{4.8E+03}^{1.7E+04}$ & $0.81_{0.46}^{0.96}$ & 84 \\[.2cm]
    CZER30 & $-0.32_{-0.45}^{-0.04}$ & $9.56_{9.48}^{9.65}$ &
    $0.29_{0.22}^{0.33}$ & $14.08_{13.90}^{14.21}$ & $6.54_{6.03}^{6.96}$ &
    $7.2E+03_{4.3E+03}^{1.1E+04}$ & $0.79_{0.58}^{0.93}$ & 119 \\[.2cm]
    ARPM2 & $-0.33_{-0.50}^{-0.07}$ & $9.61_{9.56}^{9.69}$ &
    $0.63_{0.58}^{0.66}$ & $15.19_{15.10}^{15.30}$ & $10.91_{10.46}^{11.46}$ &
    $9.8E+03_{7.1E+03}^{1.4E+04}$ & $0.34_{0.17}^{0.57}$ & 195 \\[.2cm]
    BH4 & $-0.29_{-0.44}^{0.15}$ & $9.10_{8.88}^{9.28}$ &
    $0.34_{0.13}^{0.43}$ & $14.55_{14.22}^{15.04}$ & $8.12_{6.97}^{10.20}$ &
    $1.8E+03_{1.1E+03}^{3.3E+03}$ & $0.64_{0.32}^{0.88}$ & 66 \\[.2cm]
    F1419 & $0.02_{-0.38}^{0.15}$ & $9.62_{9.49}^{9.85}$ &
    $0.57_{0.50}^{0.65}$ & $14.82_{14.43}^{14.99}$ & $9.21_{7.71}^{9.95}$ &
    $1.1E+04_{6.8E+03}^{2.0E+04}$ & $0.63_{0.40}^{0.88}$ & 142 \\[.2cm]
    BH37 & $-0.06_{-0.41}^{0.20}$ & $8.87_{8.63}^{9.64}$ &
    $1.22_{0.80}^{1.34}$ & $12.28_{11.00}^{12.79}$ & $2.85_{1.58}^{3.61}$ &
    $2.5E+03_{1.4E+03}^{4.3E+03}$ & $0.52_{0.31}^{0.80}$ & 90 \\[.2cm]
    E9205 & $-0.12_{-0.44}^{0.13}$ & $9.78_{9.72}^{9.80}$ &
    $0.11_{0.06}^{0.17}$ & $15.52_{15.43}^{15.58}$ & $12.70_{12.22}^{13.04}$ &
    $3.3E+04_{2.7E+04}^{4.3E+04}$ & $0.74_{0.65}^{0.86}$ & 378 \\[.2cm]
    E9218 & $-0.30_{-0.33}^{-0.30}$ & $9.68_{9.68}^{9.71}$ &
    $0.24_{0.23}^{0.25}$ & $15.25_{15.23}^{15.31}$ & $11.24_{11.12}^{11.53}$ &
    $5.2E+04_{4.5E+04}^{5.9E+04}$ & $0.60_{0.52}^{0.69}$ & 721 \\[.2cm]
    SAU3 & $0.02_{-0.34}^{0.19}$ & $9.81_{9.51}^{9.91}$ &
    $0.71_{0.64}^{0.78}$ & $13.94_{13.77}^{14.30}$ & $6.12_{5.68}^{7.26}$ &
    $1.4E+04_{8.3E+03}^{2.0E+04}$ & $0.86_{0.67}^{0.96}$ & 146 \\[.2cm]
    KRON39 & $-0.11_{-0.43}^{0.15}$ & $9.45_{9.34}^{9.67}$ &
    $0.78_{0.69}^{0.86}$ & $15.57_{15.25}^{15.80}$ & $13.03_{11.22}^{14.44}$ &
    $1.3E+04_{6.4E+03}^{2.6E+04}$ & $0.75_{0.41}^{0.94}$ & 55 \\[.2cm]
    E9308 & $-0.32_{-0.52}^{0.02}$ & $9.91_{9.54}^{10.05}$ &
    $0.67_{0.60}^{0.72}$ & $15.61_{15.38}^{15.85}$ & $13.25_{11.91}^{14.80}$ &
    $3.4E+04_{1.4E+04}^{6.9E+04}$ & $0.68_{0.32}^{0.89}$ & 60 \\[.2cm]
    BH144 & $-0.53_{-0.55}^{-0.48}$ & $9.02_{8.99}^{9.03}$ &
    $0.83_{0.80}^{0.84}$ & $15.01_{14.95}^{15.09}$ & $10.06_{9.79}^{10.41}$ &
    $6.9E+03_{6.1E+03}^{7.6E+03}$ & $0.32_{0.25}^{0.41}$ & 307 \\[.2cm]
    BH176 & $0.15_{-0.03}^{0.26}$ & $9.70_{9.62}^{9.71}$ &
    $0.52_{0.49}^{0.58}$ & $16.31_{16.24}^{16.39}$ & $18.27_{17.72}^{18.96}$ &
    $1.7E+05_{1.3E+05}^{1.9E+05}$ & $0.48_{0.35}^{0.61}$ & 277 \\[.2cm]
    KRON31 & $-0.37_{-0.49}^{0.13}$ & $9.00_{8.92}^{9.06}$ &
    $1.32_{1.22}^{1.35}$ & $14.40_{14.23}^{14.74}$ & $7.57_{7.02}^{8.87}$ &
    $8.3E+03_{5.7E+03}^{1.2E+04}$ & $0.80_{0.65}^{0.93}$ & 133 \\[.2cm]
    SAU6 & $-0.11_{-0.40}^{0.15}$ & $9.14_{9.05}^{9.44}$ &
    $0.94_{0.85}^{1.04}$ & $14.82_{14.26}^{15.10}$ & $9.19_{7.10}^{10.48}$ &
    $5.2E+03_{3.5E+03}^{8.2E+03}$ & $0.52_{0.33}^{0.74}$ & 129 \\[.2cm]
    BER56 & $-0.34_{-0.35}^{-0.34}$ & $9.72_{9.71}^{9.72}$ &
    $0.51_{0.50}^{0.51}$ & $15.23_{15.18}^{15.25}$ & $11.12_{10.87}^{11.23}$ &
    $6.1E+04_{5.3E+04}^{6.7E+04}$ & $0.70_{0.60}^{0.75}$ & 843 \\[.2cm]
    BER102 & $-0.17_{-0.45}^{0.07}$ & $9.69_{9.63}^{9.84}$ &
    $0.45_{0.39}^{0.51}$ & $14.33_{14.14}^{14.58}$ & $7.35_{6.72}^{8.22}$ &
    $5.8E+03_{4.2E+03}^{9.2E+03}$ & $0.55_{0.36}^{0.75}$ & 156 \\[.2cm]
  \hline
  \end{tabular}
  \end{table*}

  Table~\ref{tab:results} shows the fundamental parameters along with their
  uncertainties estimated by \texttt{ASteCA}. The Bayesian inference process
  was allowed to run for enough steps to achieve convergence.

\begin{figure*}
   \resizebox{\hsize}{!}{\includegraphics[]{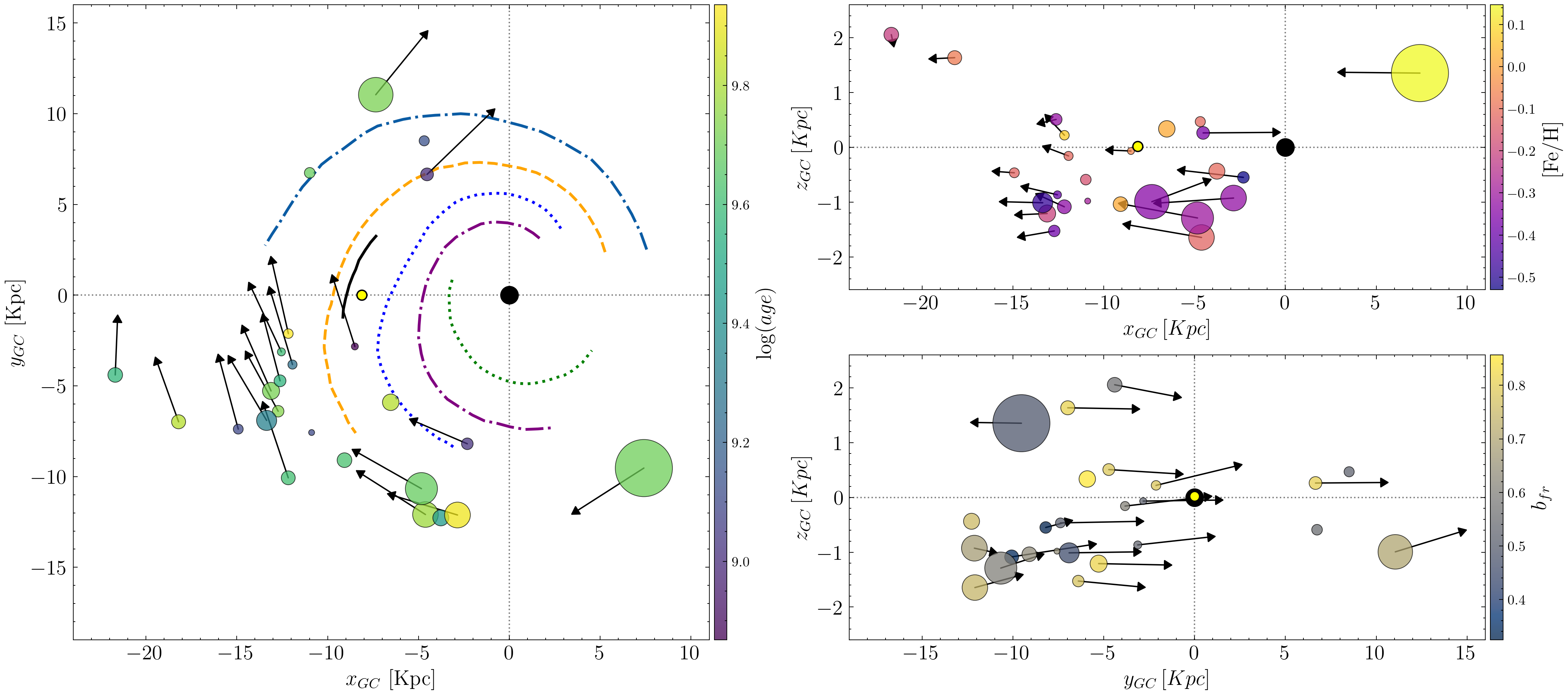}}
   \caption{Same as Fig.~\ref{fig:MWmap} but showing the positions given by our
   analysis with \texttt{ASteCA}. The velocity vectors are drawn for those
   clusters with available radial velocities. The length of the vectors are
   proportional to the velocity modules in each 2D projection. Sizes follow
   masses and colors follow ages, metallicities, and binary fractions, for the
   left, top right, and bottom right plots, respectively.}
   \label{fig:MWmap_vectors}
  \end{figure*}

  In Fig.~\ref{fig:MWmap_vectors} we show how the map of the Galaxy shown
  previously in Fig.~\ref{fig:MWmap} looks, but with the distance parameter
  values found in this work. The arrows represent the velocity vectors for all
  the clusters with available radial velocity. Sizes correspond to the estimated
  masses, and colors follow the distribution of ages, metallicities, and binary
  fraction as shown in the colorbars to the right of each plot. The values used
  to construct this figure are presented in Table~\ref{tab:velocities}.
  It is worth noting that only about half of the clusters are truly beyond the 9
  kpc ($\sim$14.8 mag) limit originally used to perform the selection from the
  published databases.
  The cluster vd Bergh-Hagen 176, located in the 4th Quadrant
  in Fig.~\ref{fig:MWmap_vectors}, turns out to be the most distant open cluster
  catalogued to date with a heliocentric distance greater than 18 kpc. Its
  status as a bonafide open cluster is nonetheless still questioned; a more
  detailed discussion is presented in Appendix~\ref{app:indiv_clusters}.

\begin{figure*}
   \resizebox{\hsize}{!}{\includegraphics[]{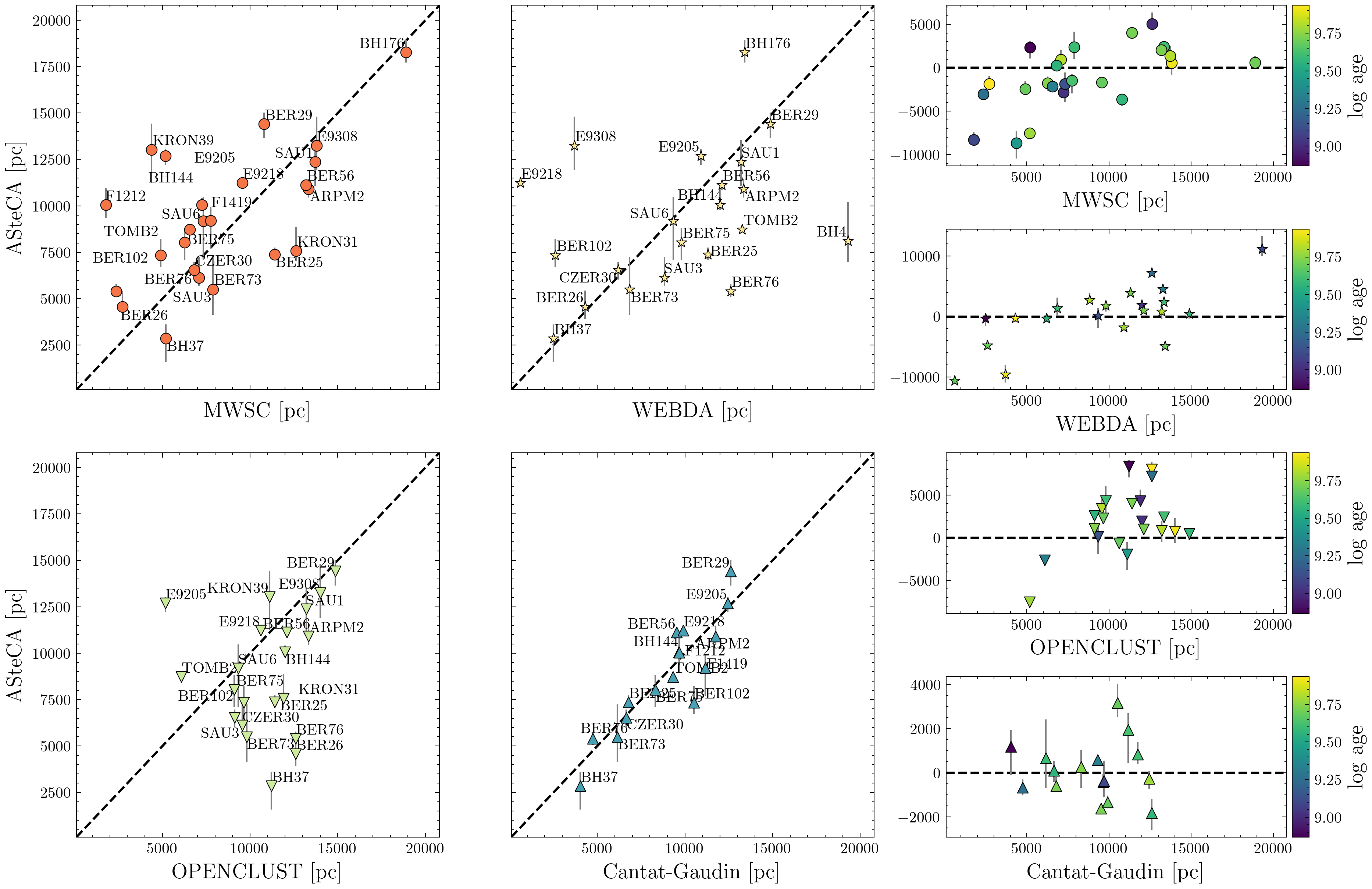}}
   \caption{\texttt{ASteCA} versus database distances. The plots to the right
   stacked vertically are the \texttt{ASteCA} distances versus the differences
   in the sense (\texttt{ASteCA} - database). Clusters are colored according to
   the $\log\,age$ assigned by \texttt{ASteCA}.}
   \label{fig:distances}
  \end{figure*}

  In a recent study~\citep{Anders_2022} per-star parameters such as distance,
  extinction, metallicity, and age were estimated. Comparing the results
  from this analysis with those from \cite{Cantat_2020}, the authors find
  differences in the distance values larger than 3 kpc for clusters located at
  6 kpc or more from the Sun. We find even larger discrepancies between our
  analysis and those taken from the four databases. As shown in
  Fig.~\ref{fig:distances}, all but the CG20 database show differences of up
  to 10 kpc for clusters spanning the full distance range. The CG20
  database, the one with the better overall match to our values, only
  shows differences larger than 2 kpc for clusters located beyond
  $\sim10$ kpc from the Sun. Taking the uncertainties of both estimates
  into account, these differences are expected; particularly for such distant
  clusters.
  % It is thus clear that even the database with the closest fit
  % contains substantial disagreements with the distance values estimated by 
  % \texttt{ASteCA}.
  We see no evident trend that correlates the differences in the distance with
  the ages (used to color the markers in the right plots of
  Fig.~\ref{fig:distances}).\\

  To further investigate the various ways to estimate the distance, we performed
  two more analyses. First, we re-run \texttt{ASteCA} for all the
  clusters using four different combinations of settings for the metallicity and
  binary fraction parameters. We chose these two parameters because in
  isochrone-fit analyses they are usually either fixed (e.g., the metallicity is
  set to solar) or neglected altogether (e.g., the binary fraction).
  Second, we compared the distances estimated in this work with those obtained
  via parallax analysis using three different methods: \texttt{ASteCA}'s own
  Bayesian inference estimation~\citep[described in][]{Perren_2020}, the
  distance inferred by the Kalkayotl package~\citep{Kalkayotl}, and the median
  of a simple inversion of the parallax values of the selected members.
  Parallax values were previously corrected using the method described
  in~\cite{Lindegren_2021}\footnote{ Analytical functions to compute the
  expected parallax zero-point as a function of ecliptic latitude, magnitude and
  colour for any Gaia (e)DR3
  source: \url{https://gitlab.com/icc-ub/public/gaiadr3_zeropoint}}.

\begin{figure}
   \resizebox{\hsize}{!}{\includegraphics[]{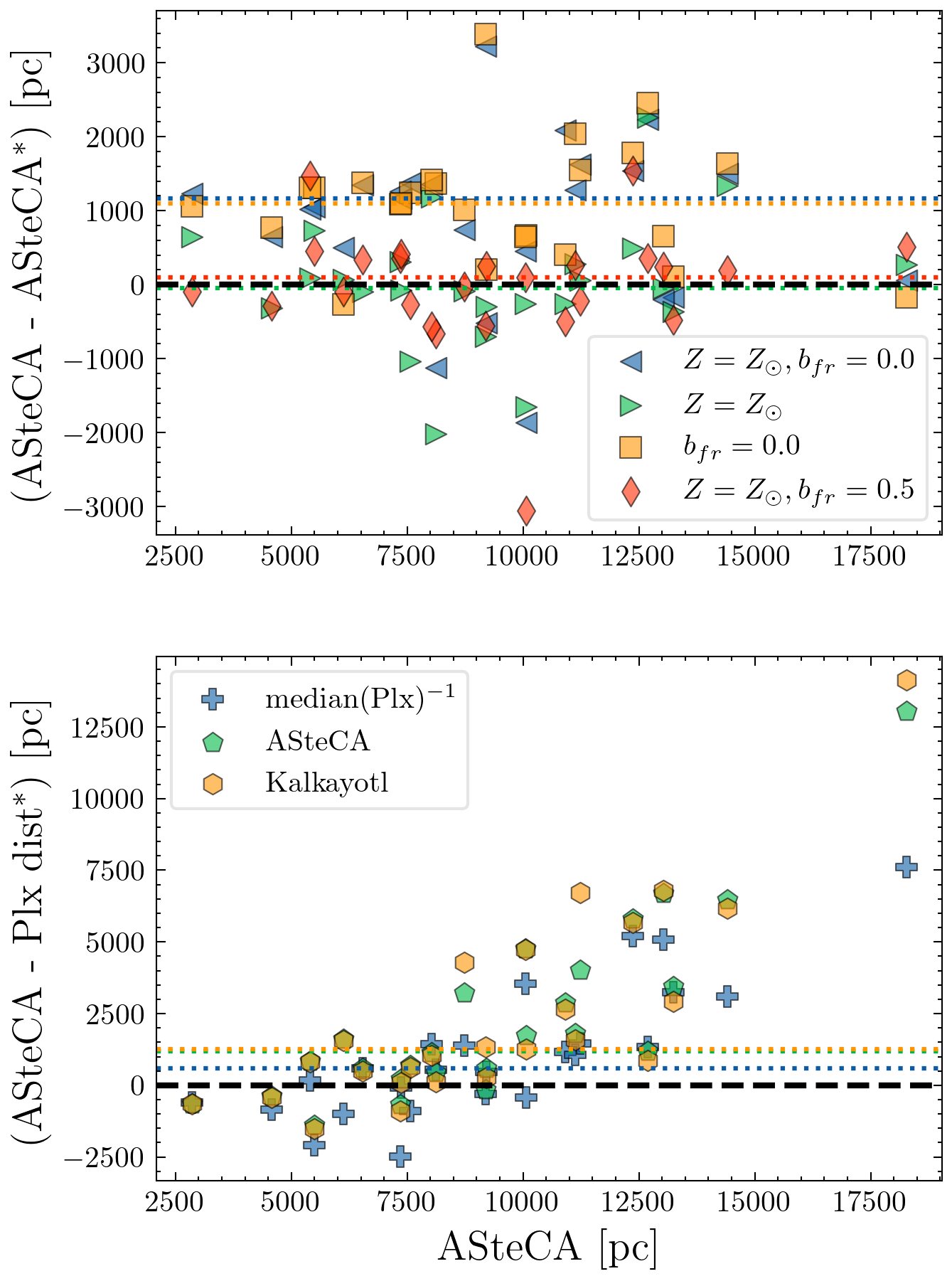}}
   \caption{Top: distances estimated with the main run of \texttt{ASteCA}
   compared to four different runs where the metallicity and binary fraction
   were wither fixed or allowed to be fitted.
   Bottom: main run distances versus the difference with the distances
   estimated using parallax values and three different methods.
   In both plots the abscissa are the main run \texttt{ASteCA} distance values
   and the ordinate shows the difference in the sense (\texttt{ASteCA} - 
   \texttt{ASteCA}$^*$), where the asterisk represents either of the four runs
   from the top plot or either method from the bottom plot.
   In both plots the black dashed line marks zero difference, and the colored
   dotted lines mark the median differences for each run.}
   \label{fig:dist_comparisions}
  \end{figure}

  The results of these two extra analysis can be seen in
  Fig.~\ref{fig:dist_comparisions} compared to the main \texttt{ASteCA} run,
  i.e., the one whose estimated parameter values are shown in
  Table~\ref{tab:results}.
  In the top plot we show the variation in the distance estimates between our
  main \texttt{ASteCA} run and four different runs: metallicity fixed to solar
  and binary fraction fixed to 0 ($Z=Z_{\odot},b_{fr}=0.0$; blue left
  facing triangles), metallicity fixed to solar and binary fraction as a free
  parameter ($Z=Z_{\odot}$; green right facing triangles), metallicity
  as a free parameter and binary fraction fixed to 0 ($b_{fr}=0.0$; orange
  squares), and metallicity fixed to solar and binary fraction fixed to
  0.5 ($Z=Z_{\odot},b_{fr}=0.5$; red diamonds), where 50\% is
  chosen to be a typical estimate for binary fraction in open
  clusters~\citep{vonHippel_2005}.
  The median difference with the main \texttt{ASteCA} run is largest when the
  binary fraction is fixed to 0.0 ($\sim$1100 pc), lower when we fix this
  parameter to 0.5 ($\sim$100 pc), and lowest when it is allowed to vary 
  ($\sim$50 pc). This is another indicator that a proper binary fraction fit is
  of utmost importance for a correct  estimation of the cluster's fundamental
  parameters, particularly for the distance. Even when the binary fraction is
  free, fixing just the metallicity to solar values can have a non
  negligible impact on the estimated distances as shown in
  Fig.~\ref{fig:dist_comparisions} (green right facing triangles).
  % $Z=Z_{\odot},b_{fr}=0.0$: 1166
  % $Z=Z_{\odot}$: -49
  % $b_{fr}=0.0$: 1098
  % $Z=Z_{\odot},b_{fr}=0.5$: 102

  The bottom plot in Fig.~\ref{fig:dist_comparisions} shows the parallax values
  analysis. Here the distance estimates obtained by \texttt{ASteCA} processing
  the Gaia EDR3 photometry are compared to three methods to estimate distances
  using parallaxes: \texttt{ASteCA}'s own Bayesian inference, the Kalkayotl
  package estimate, and the inversion of the median of the selected member's
  parallaxes. It is clear to see that a trend arises where the most distant
  clusters have their distances enormously underestimated by any of the
  parallax-based methods. This is expected, as the parallax values of the most
  distant clusters are associated to very large uncertainties and are also
  heavily affected by noise from non-removed field stars that mostly contaminate
  the lower mass region.
  It is surprising to see that the naive approach of inverting the median of
  the member's parallaxes is the method that more closely approximates the
  photometric distances estimated by \texttt{ASteCA}: the mean difference is
  only $\sim$600 pc, where the other two methods show median differences more
  that twice as large ($\sim$1200 pc).
  % median(Plx)$^{-1}$: 581
  % ASteCA: 1183
  % Kalkayotl: 1253
  \\

\begin{figure}
   \resizebox{\hsize}{!}{\includegraphics[]{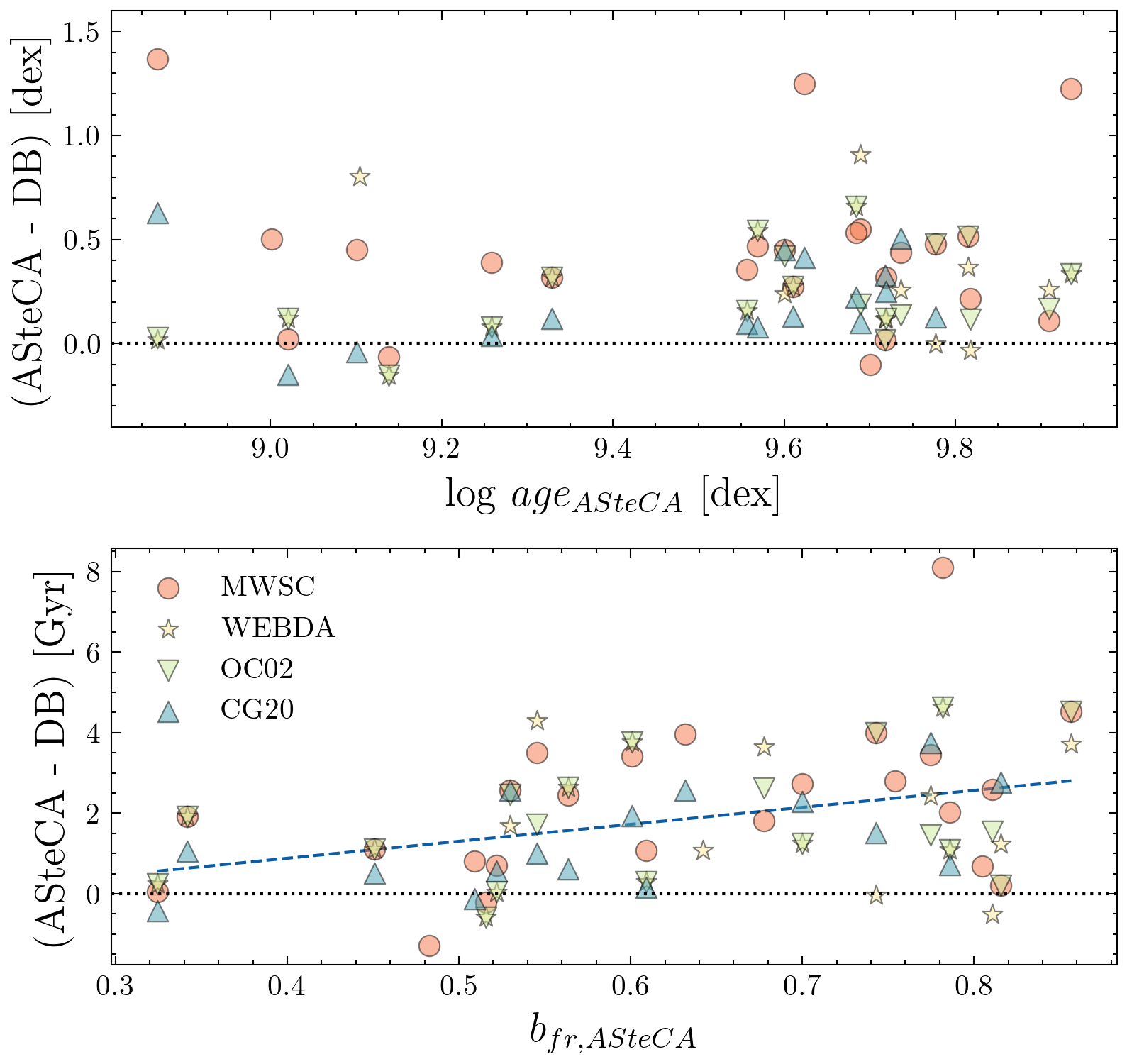}}
   \caption{Top: differences in logarithmic ages between \texttt{ASteCA} and
   the databases in the sense (\texttt{ASteCA} - database). The cluster
   Kronberger 39 is left out of the plot to improve visibility.
   Bottom: linear age differences versus binarity fraction assigned by
   \texttt{ASteCA}. The blue dashed line is the regression trend.}
   \label{fig:ages}
  \end{figure}

  All the analyzed clusters are rather old, with the youngest one (vd
  Bergh-Hagen 37) assigned an age of $\sim0.7$ Gyr; although notice the very
  large uncertainty associated to it.
  The comparison between our age estimation and those from
  the four databases are shown in Fig.~\ref{fig:ages}. The top plot
  shows that \texttt{ASteCA} systematically assigns ages that are larger on
  average than those from the databases.
  A logarithmic difference of $\sim$0.23 dex (the average value for
  all the catalogs) translates to a difference of $\sim$1.5 Gyr for
  an age of 3 Gyr, which is a reasonable uncertainty given the complexities
  associated to the clusters under investigation. The catalog with the
  smallest logarithmic difference is CG20 with a median of 0.11 dex, again
  displaying the best match to the values given by \texttt{ASteCA}.
  The largest age difference arises for Kronberger 39, for
  which \texttt{ASteCA} finds an age of $\log(age)=9.45$ but has an age of
  $\log(age)=6$ assigned in the MWSC database (the smallest age by far in the four
  databases).

  As can be seen in the bottom plot of Fig.~\ref{fig:ages}, there appears to be
  a slight correlation between the difference in age estimates and the
  binarity fractions. The trend shows that the larger the percentage of
  binary systems present in the cluster, the larger on average is the difference
  between the age value obtained by \texttt{ASteCA} and those in the
  databases. Such an effect can be explained by noticing that the TO in the CMD
  is pushed downwards by the presence of binaries, which are located above the
  brightest point of the main sequence of single stars. This in turn forces the
  fit to adjust towards larger ages, hence producing the systematic trend
  seen in the analysis. This result points to the importance of taking binary
  systems into account when performing stellar clusters' parameters estimations.
  \\

  %
  % Metallicity
  %

  % Donor et al 2020 (Vizier)
  % Berkeley 29: -0.49 0.03 (3); Czernik 30: -0.4  0.01 (2); Saurer 1: -0.42
  % 0.01 (1)
  %
  % \cite{Netopil_2022}
  % Berkeley 25   -0.2  0.04  6
  % Berkeley 29   -0.49 0.03  3 <-- Donor
  % Berkeley 73   -0.23 0.06  2
  % Czernik 30    -0.4  0.01  2 <-- Donor
  % Saurer 1      -0.42 0.01  1 <-- Donor
  % Tombaugh 2    -0.3  0.08  17
  %
  % \cite{Spina_2021}:
  % BER29: -0.48 0.013 (1)
  % BER73: -0.319 0.058 (1)
  % CZER30: -0.396 0.006 (2) 
  %
  % \cite{Carraro_2007_oldOC}
  % Berkeley 25: -0.20 ± 0.05 (4)
  % Berkeley 73: -0.22 ± 0.10 (2)
  %
  % \cite{Frinchaboy_2006}
  % Berkeley 29: -0.44
  % Saurer 1: -0.38
  % vd BH 176: --
  %
  % \cite{Carraro_2004}
  % BER 29: -0.44 ± 0.18
  % SAU1:   -0.38 ± 0.14 
  %
  % \cite{Frinchaboy_2004}
  % BER29: -0.62 (8)
  % SAU1:  -0.49 (2)
  % BH144: -0.51 (2)
  \begin{table}
  \caption{Six cluster from our sample whose [Fe/H] metal content was also
  analyzed in recent works. In parenthesis, the number of stars used to estimate
  each value.}
  \label{tab:metal}
  \centering
  \begin{tabular}{lcccc}
  \hline\hline
  Cluster & \texttt{ASteCA} & Donor & Netopil & Spina \\
  \hline %\\[.2cm]
   BER25  & -0.20  & --         & -0.20 (6)  & -- \\
   BER29  & -0.21  & -0.49 (3)  & --         & -0.48 (1)  \\
   BER73  & -0.41  & --         & -0.23 (2)  & -0.319 (1)  \\
   CZER30 & -0.32  & -0.40 (2)  & --         & -0.396 (2)  \\
   SAU1   & -0.08  & -0.42 (1)  & --         & --  \\
   TOMB2  & -0.48  & --         & -0.30 (17) & -- \\
  \hline
  \end{tabular}
  \end{table}

\begin{figure}
   \resizebox{\hsize}{!}{\includegraphics[]{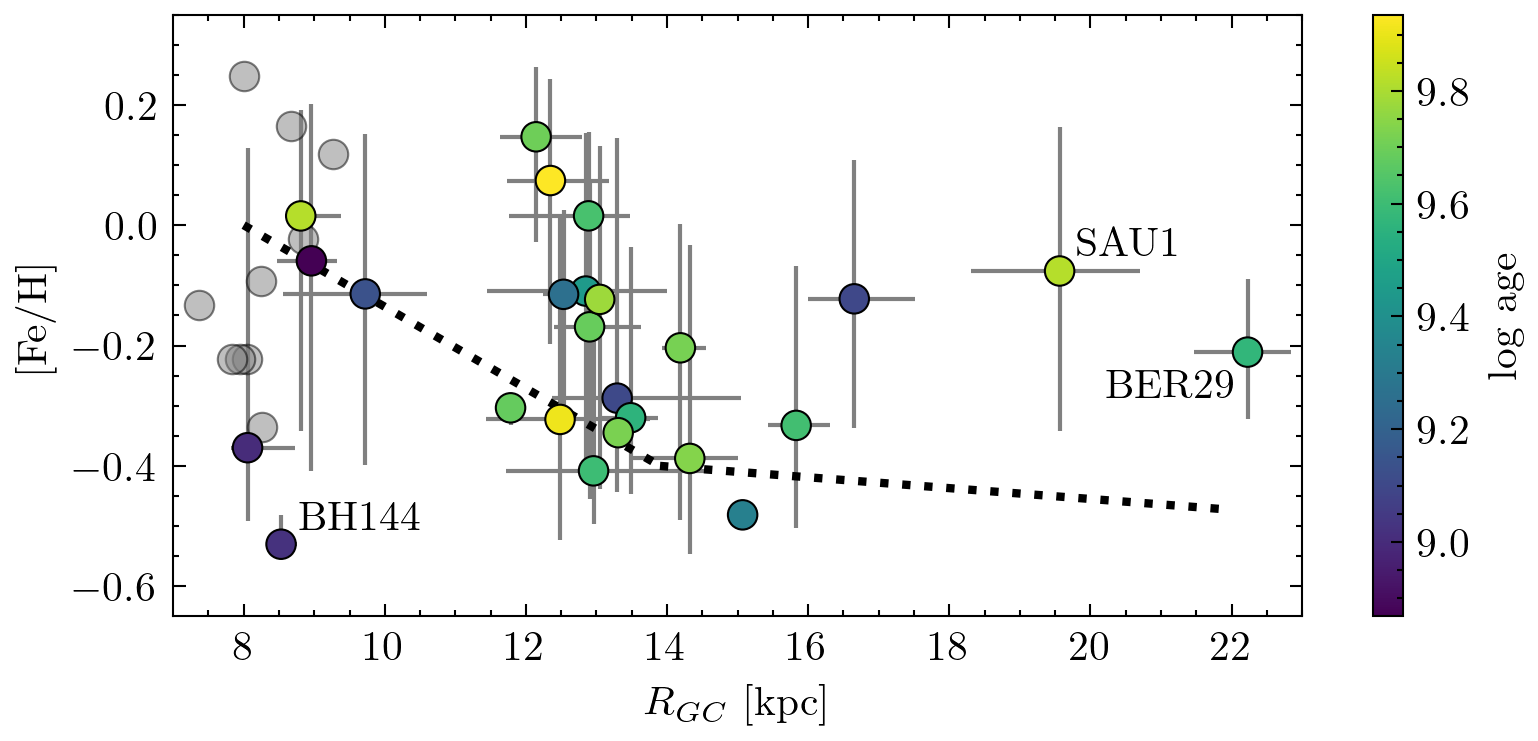}}
   \caption{Metallicity gradient for the set of twenty-five analyzed clusters.
   Points are colored according to the $\log(age)$. Grey vertical lines are the
   16th and 84th percentiles. The dotted line is the broken relation from 
   \citet[][Fig 7]{Donor_2020}. The grey dots are the ten verified clusters
   from~\cite{Perren_2020}.}
   \label{fig:met_gradient}
  \end{figure}

  The metal content of a cluster is the hardest parameter to estimate
  photometrically, which is why it is usually fixed to solar value in these
  type of analysis.
  We found six clusters from our sample that were also investigated
  spectroscopically in very recent works: Berkeley 25, Berkeley 29,
  Berkeley 73, Czernik 30, Saurer 1, and Tombaugh 2; studied
  in~\cite{Donor_2020}, \cite{Netopil_2022}, and \cite{Spina_2021}. The
  abundances are shown in Table~\ref{tab:metal} along with the \texttt{ASteCA}
  estimates. Uncertainties are around 0.02, 0.06, and 0.04 dex for Donor et
  al., Netopil et al., and Spina et al., respectively.
  The uncertainties associated to the \texttt{ASteCA} values are substantially
  larger, averaging 0.2 dex (see Table~\ref{tab:results}).
  In Fig.~\ref{fig:met_gradient} we show the metallicity versus
  galactocentric distance ($R_{GC}$) distribution for the clusters in our
  sample, plus ten verified clusters from~\cite{Perren_2020}.
  This distribution (also called radial metallicity distribution or
  metallicity gradient) is a key tracer of the Galaxy's chemical evolution.
  Open clusters have been used as a tool to investigate this relation for several
  decades~\citep{Janes_1979}. The gradient is usually taken to be around -0.05
  dex kpc$^{-1}$ for the inner clusters, with a break beyond $R_{GC}\approx10$
  kpc into a shallower slope~\citep{Donor_2020}. In~\cite{Donor_2018} it was
  previously shown that the metallicity gradient is also (expectedly) highly
  dependent on the database used fix the distances, varying as much as 40\%
  depending on the used database. This result was confirmed in~
  \cite{Donor_2020}, where a database-dependent variation of of 15\% was found.
  % from Donor et al (2020)
  % Donor et al. (2018) (henceforth OCCAMII) showed that this gradient could
  % change by as much as 40%, depending on which distance catalog was used
  %

  % Although we have a somewhat small sample, we
  % present here the results to highlight the difficulties that arise when
  % estimating the metal content from photometric data alone.

  We see in Fig.~\ref{fig:met_gradient} that \texttt{ASteCA} assigns on
  average a slightly larger metallicity ($\sim$0.06 dex) than that
  expected for clusters located below $R_{GC}\approx14$ kpc.
  % There is a visible concentration of \textbf{seven} clusters around $R_
  % {GC}\sim13$ kpc grouped around [Fe/H]$\approx0.0$ dex, where the Donor et al.
  % linear relation predicts a value of [Fe/H]$\approx-0.3$ dex.
  The case of vd Bergh-Hagen 144 is interesting because its estimated
  metallicity of [Fe/H]$=-0.53_{-0.48}^{-0.55}$ is well below the expected
  solar value at that distance ($R_{GC}\sim$8.5 kpc). There are two other
  articles where a similar markedly sub-solar metallicity was found
  for this cluster: \cite{Frinchaboy_2004} and \cite{Fragkou_2019}. In these
  studies the reported metallicity values are [Fe/H]=-0.51$\pm$0.3 
  (spectroscopic metallicity from 2 stars) and [Fe/H]$\approx$-0.40 (photometric
  estimate)\footnote{Transformed from the fitted Z=0.006 assuming
  $Z_{\odot}=0.0152$}, for Frinchaboy et al. and Fragkou et al. respectively.
  Fragkou et al. assigned a distance of 12$\pm$0.5 kpc, $\sim4$ kpc
  larger than the one found by \texttt{ASteCA}, while in~\cite{Frinchaboy_2004b}
  the estimated distance is 9.35 kpc which is a much closer value to ours.

  For the seven clusters beyond this galactocentric distance the
  difference with \texttt{ASteCA} is larger, where our code assigns abundances
  on average 0.20 dex above the Donor et al. gradient.
  Saurer 1 is the cluster with the largest value in this group, with
  an abundance assigned by \texttt{ASteCA} of [Fe/H]=$-0.08_{-0.34}^{0.16}$
  which conflicts with the value $\sim-0.42$ dex expected for its
  galactocentric distance.
  %
  % the value expected for its galactocentric distance of $\sim19.7$ kpc.
  %
  There are several articles where this clusters was assigned a 
  markedly sub-solar [Fe/H] value: -0.27~\citep{Carraro_2003},
  -0.38$\pm$0.14~\citep{Carraro_2004},
  -0.50~\citep{Frinchaboy_2004b}, -0.38~\citep{Frinchaboy_2006},
  -0.42$\pm$0.01~\citep{Donor_2020}. The distances given to Saurer 1 in these
  studies are located in the range [12, 13.2] kpc, a reasonable match
  for the distance estimated by \texttt{ASteCA} of $12.4_{11.1}^{13.5}$ kpc. It
  is thus clear that \texttt{ASteCA} has overestimated the metal
  content for this cluster. Saurer 1 is the third oldest cluster in our sample 
  ($\sim6.6$ Gyr) and one of the most distant from the Sun, which results in
  less than a  full magnitude visible below the TO with a total of only 84
  members present in the CMD. This is very likely the reason
  for the large difference in the abundance estimated by \texttt{ASteCA}
  versus the one predicted by the radial metallicity trend and the
  spectroscopic analyses.
  % SAU1 distance:
  % ASteCA: 12.5
  % Carraro 2003: 13.2
  % Carraro 2004: 13.2
  % Frinchaboy_2004b: 11.97
  % Frinchaboy_2006: 13.1
  % Donor_2020: ??
  The clusters analyzed in~\cite{Perren_2020} on the other hand (shown as grey
  circles in the plot) display a much more balanced distribution around the 
  [Fe/H]$\approx0.0$ dex expected value for their galactocentric distance of
  $\sim9$ kpc.
  The abundances presented here should therefore be taken with
  caution. It is always preferable to refer to spectroscopical estimates
  whenever available, particularly when dealing with very distant, old,
  scarcely populated, and/or heavily contaminated clusters.\\

  The assigned binary fractions range from 32\% up to $\sim86$\%, 
  with a mean value of 63\% for the entire sample. Although this value
  is not that far off from the typically value expected for open clusters (50\%
  as stated previously), it is a bit large. On the other hand, the
  uncertainties are also rather large and the binary fractions assigned to
  most of the clusters drop below 50\% within their 1-sigma range.
  In~\cite{Sollima_2010} the binary fraction within the core radius of five
  clusters is estimated. The authors find values in the range 35\%-70\% with a
  combined mean value of 56\%, somewhat similar to ours. We did not estimate
  core binary fractions, but we did employ radii about 50\% smaller than the
  cluster's tidal radii, so the large binarity found could be related to this.
  In any case, as these are rather distant and old clusters most of which have
  a very small portion of their sequence observed, these values should also be
  used with caution.\\

  The total estimated masses are found in the range [2000, 60000] M$_{\odot}$
  with the exception of vd Bergh-Hagen 176, for which a much
  larger mass of $\sim$170000 M$_{\odot}$ is given. This is a large mass
  value for an open cluster which would suggest that this object is closer to
  being classified as a young globular cluster.
  It is worth noting that these are lower limit mass
  estimates since \texttt{ASteCA} does not take into account the
  experienced dynamical mass loss, which can be significant for old stellar
  clusters~\citep{Martinez_2017}.\\

% =============================================================================
\section{Conclusions}
 \label{sec:conclusions}

  Taking advantage of the precise photometry and proper motions from the
  most recent Gaia data release~\citep{Gaia_EDR3}, the fundamental parameters
  of the twenty-five most distant catalogued clusters ($>9$ kpc) have been
  reassessed using pyUPMASK and \texttt{ASteCA}.
  The results for the fundamental parameters metallicity, age, distance,
  extinction, total mass, and binary fraction, are shown in 
  Table~\ref{tab:results}. In this table we can see that these are
  rather old clusters: with the exception of just two (vd Bergh-Hagen 37 and
  Kronberger 31) the remaining ones are all older than 1 Gyr.
  Only thirteen clusters out of twenty-five turn out to be at a distance larger
  than 9 kpc from the Sun, thus reducing the number of clusters that fit the
  minimum distance criteria by almost half.

  Regarding the distribution in the galactic plane and the galactocentric
  distance, we see that fourteen clusters are placed in the third galactic
  quadrant with ten out of these having negative latitudes thus located below
  the formal galactic equator. The remaining four clusters that are above the
  galactic plane are Berkeley 26 ($Z\approx$0.2 kpc), Saurer 1
  ($Z\approx$1.6 kpc), Czernik 30 ($Z\approx$0.5 kpc), and Berkeley
  29 ($Z\approx$2 kpc). The cluster with the largest galactocentric distance,
  Berkeley 29, is also the cluster with the largest vertical distance. This
  cluster is on its course to cross the galactic disk, as shown by its velocity
  vector seen in Fig.~\ref{fig:MWmap_vectors}. Maximum vertical distances
  of $\sim$1.8 kpc for Berkeley 29 and $\sim$1.6 kpc for Saurer 1 are estimated
  in~\cite{Gaia_Collaboration_2021}. These are in reasonable agreement with the
  vertical distances obtained here, meaning that both clusters are currently at
  their maximum height above the galactic plane.

  Despite some bias effect~\citep[e.g. lower dust absorption, particularly
  along the Fitzgerald window][]{Fitzgerald_1968}, it appears that a large
  number of clusters in the third quadrant of the Galaxy follow the warp defined
  by the diffuse blue population~\citep{Carraro_2005_detection,Moitinho_2006},
  whose maximum height above the galactic equator takes place at about
  (lat=-8$^{\circ}$, lon=240$^{\circ}$.
  The four databases list fourteen clusters with galactocentric distances larger
  than 15 kpc (in either one of them), the assumed limit for the galactic disk
  radius~\citep[see][and references therein]{Carraro_2010}.
  One of the most relevant result emerging from our new distance estimation is
  that five clusters were confirmed to be located beyond this value.
  These are: Tombaugh 2 ($R_{GC}\approx$15.1 kpc), Arp-Madore 2
  ($R_{GC}\approx$15.8 kpc), FSR 1212 ($R_{GC}\approx$16.7 kpc), Saurer 1
  ($R_{GC}\approx$19.6 kpc), and Berkeley 29 ($R_{GC}\approx$22.2 kpc). These
  values are in line with recent findings where evidences of population more than
  15 kpc away from the galactic center were presented~\citep[][and references
  therein]{Liu_2017,Lopez_2018}.
  In this work we are reporting distant open clusters (older tan 1.2 Gyr)
  instead of single stars, as shown in most of the papers referred above.
  A recent review of the spatial distribution of star clusters and
  the impact of the subsequent releases of Gaia data on the topic can be found
  in~\cite{Cantat-Gaudin_2022}.\\

  When comparing the results given by our analysis with \texttt{ASteCA} with
  those present in the MWSC, WEBDA, and OC02 databases, we clearly
  see substantial disagreements in age and distance (the fundamental
  parameters available in these catalogs).
  The best overall agreement in distances, the main objective of this article,
  is found with the database presented in CG20. For the sixteen clusters in
  common with this work the differences range from -2 to +3 kpc, without any
  apparent dependency on the ages. Within the limits of the uncertainties
  associated to the distance parameter, we can say that the agreement with
  CG20 is good.
  The differences with other databases are substantially larger, spanning a
  range from -10 to +10 kpc, and are present in the case of MWSC and WEBDA for
  clusters whose catalogued distance is even below 5 kpc.
  The age parameter suffers also from important inconsistencies with
  the best match found again with the CG20 catalog.
  Caution is hence advised when making use of these databases for large scale
  analysis. We thus recommend choosing the CG20 database over the rest whenever
  possible.

% =============================================================================

\begin{acknowledgements}
We are grateful for the suggestions and comments given by the referee,
which greatly improved the final version of this paper.
G.I.P., M.S.P., and R.A.V. acknowledge the financial support from CONICET 
(PIP317) and the UNLP (PID-G148 project).
This work has made use of data from the European Space Agency (ESA) mission
{\it Gaia} (\url{https://www.cosmos.esa.int/gaia}), processed by the {\it Gaia}
Data Processing and Analysis Consortium (DPAC,
\url{https://www.cosmos.esa.int/web/gaia/dpac/consortium}). Funding for the DPAC
has been provided by national institutions, in particular the institutions
participating in the {\it Gaia} Multilateral Agreement.
This research has made use of the WEBDA database, operated at the Department of
Theoretical Physics and Astrophysics of the Masaryk University.
This research has made use of the VizieR catalog access tool, operated at CDS,
Strasbourg, France~\citep{Ochsenbein_2000}.
This research has made use of ``Aladin sky atlas'' developed at
CDS, Strasbourg Observatory, France~\citep{Bonnarel2000,Boch2014}.
This research has made use of NASA's Astrophysics Data System.
This research made use of the Python language v3.7.3~\citep{vanRossum_1995}
and the following packages:
NumPy\footnote{\url{http://www.numpy.org/}}~\citep{vanDerWalt_2011};
SciPy\footnote{\url{http://www.scipy.org/}}~\citep{Jones_2001};
Astropy\footnote{\url{http://www.astropy.org/}}, a community-developed core
Python package for Astronomy \citep{astropy:2013, astropy:2018};
matplotlib\footnote{\url{http://matplotlib.org/}}~\citep{hunter_2007};
scikit-learn\footnote{\url{https://scikit-learn.org/}}~\citep{scikit-learn};
\texttt{ASteCA}\footnote{\url{https://github.com/asteca}};
ptemcee\footnote{\url{https://github.com/willvousden/ptemcee}};
Kalkayotl\footnote{\url{https://github.com/olivares-j/Kalkayotl}};
sfdmap\footnote{\url{https://github.com/kbarbary/sfdmap}}.
\end{acknowledgements}

\bibliographystyle{aa}
\bibliography{biblio} % your references Yourfile.bib

\begin{appendix}

\FloatBarrier
\section{Structure analysis}
 \label{app:struct_analysis}

\begin{figure*}
  \resizebox{\hsize}{!}{\includegraphics[]{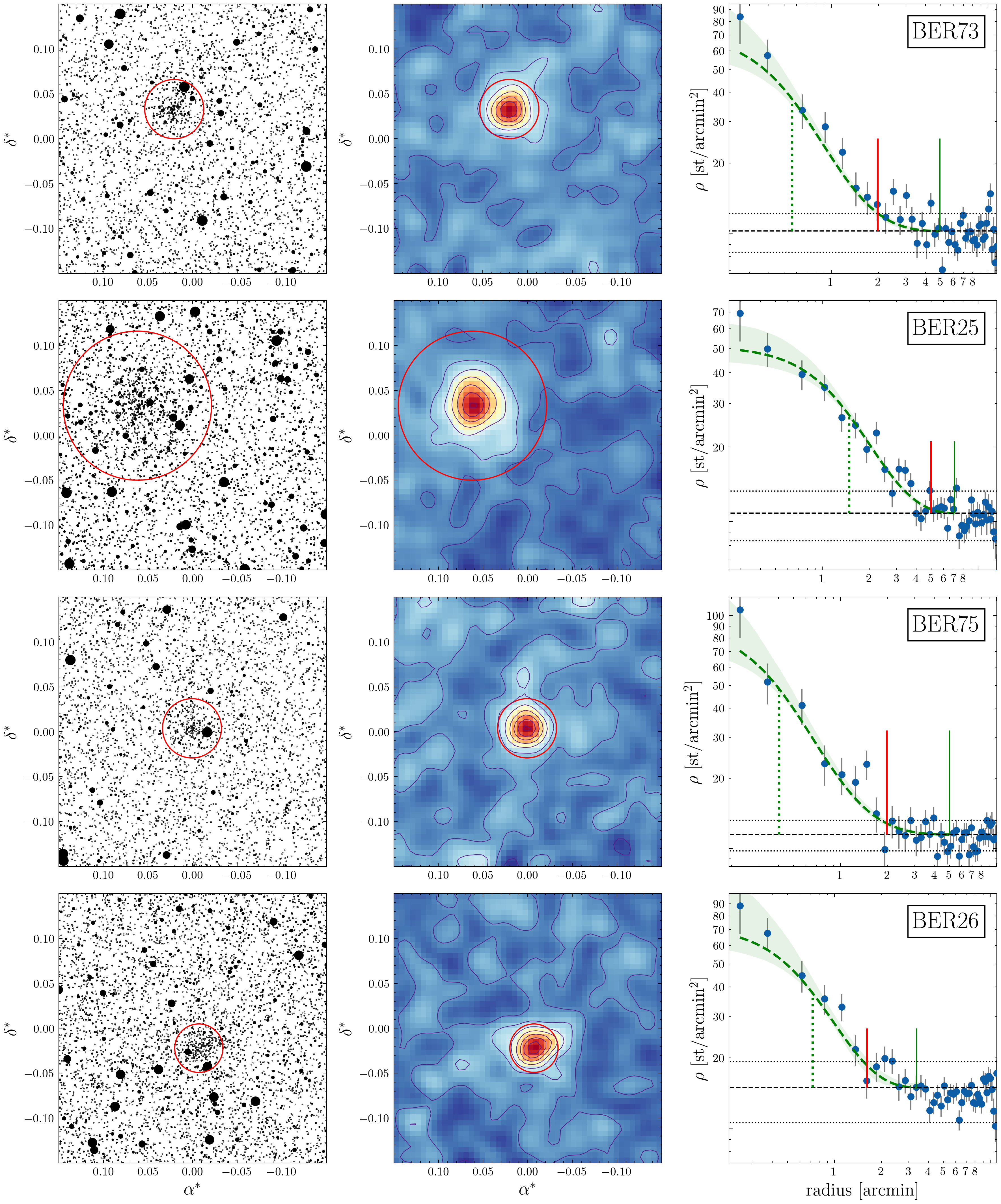}}
  \caption{Same as Fig.~\ref{fig:BER29_struct} for BER73, BER25, BER75, and BER26.}
  \label{fig:0struct}
 \end{figure*}

\begin{figure*}
  \resizebox{\hsize}{!}{\includegraphics[]{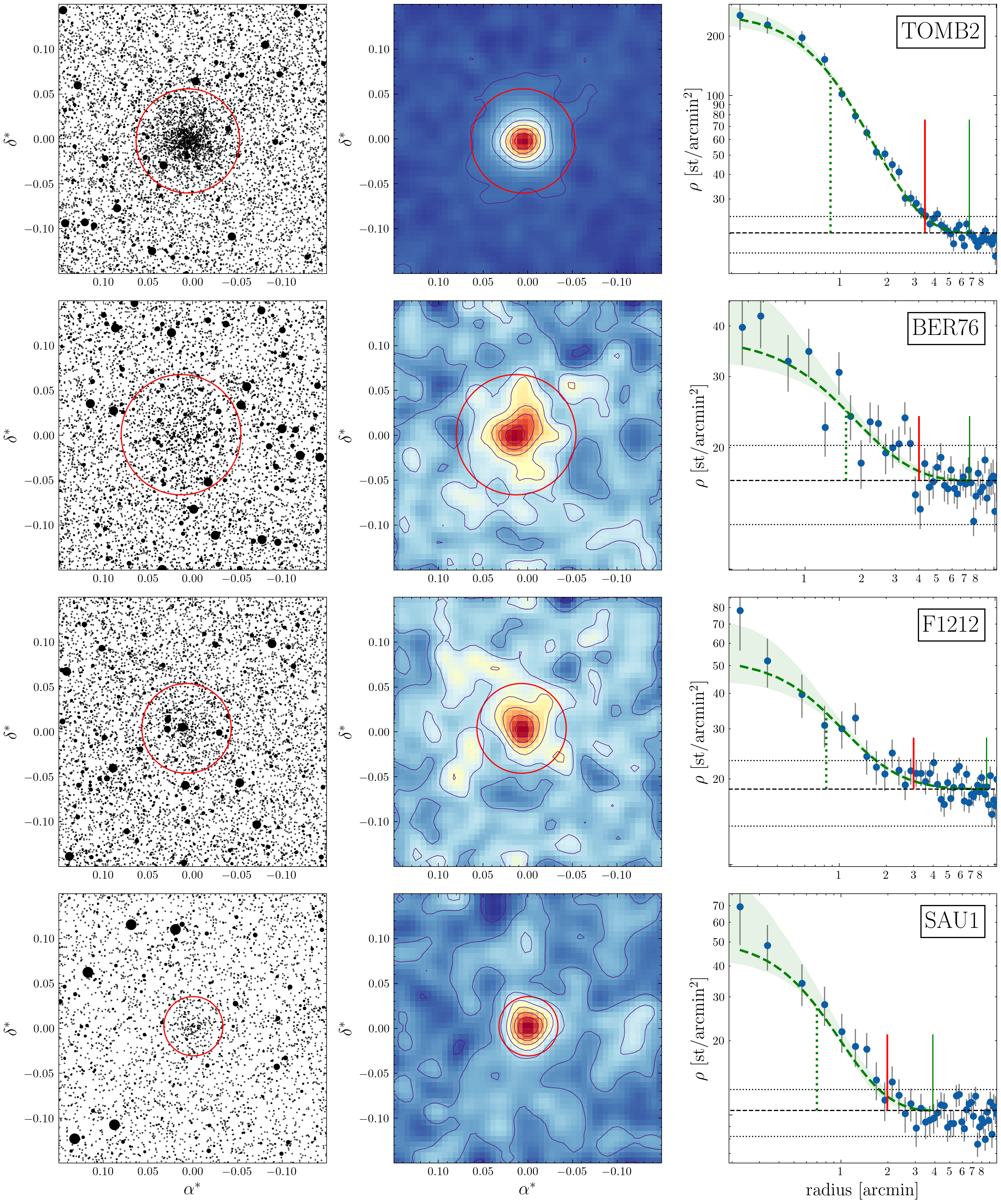}}
  \caption{Same as Fig.~\ref{fig:BER29_struct} for TOMB2, BER76, F1212, and SAU1.}
  \label{fig:4struct}
 \end{figure*}

\begin{figure*}
  \resizebox{\hsize}{!}{\includegraphics[]{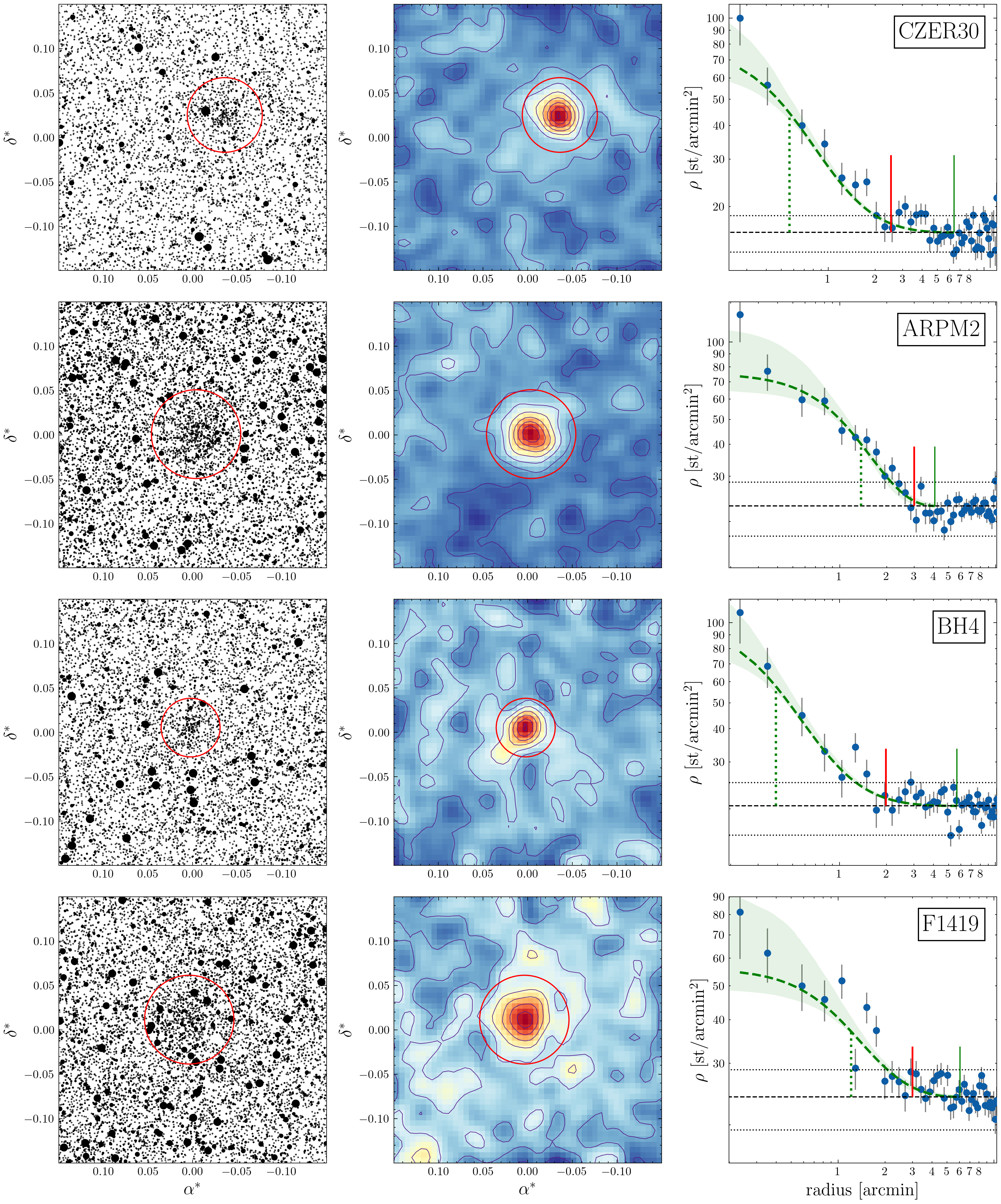}}
  \caption{Same as Fig.~\ref{fig:BER29_struct} for CZER30, ARPM2, BH4, and F1419.}
  \label{fig:8struct}
 \end{figure*}

\begin{figure*}
  \resizebox{\hsize}{!}{\includegraphics[]{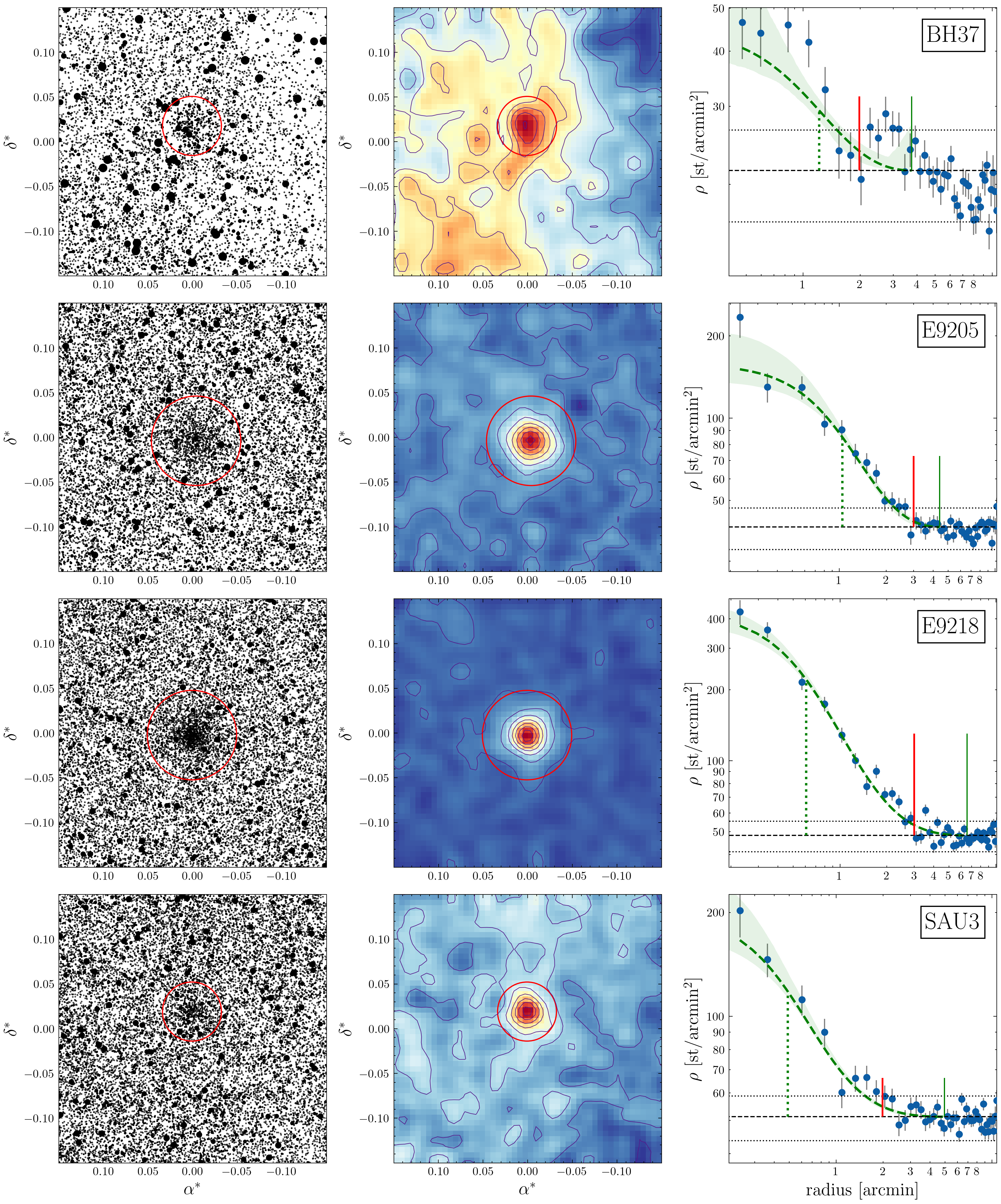}}
  \caption{Same as Fig.~\ref{fig:BER29_struct} for BH37, E9205, E9218, and SAU3.}
  \label{fig:12struct}
 \end{figure*}

\begin{figure*}
  \resizebox{\hsize}{!}{\includegraphics[]{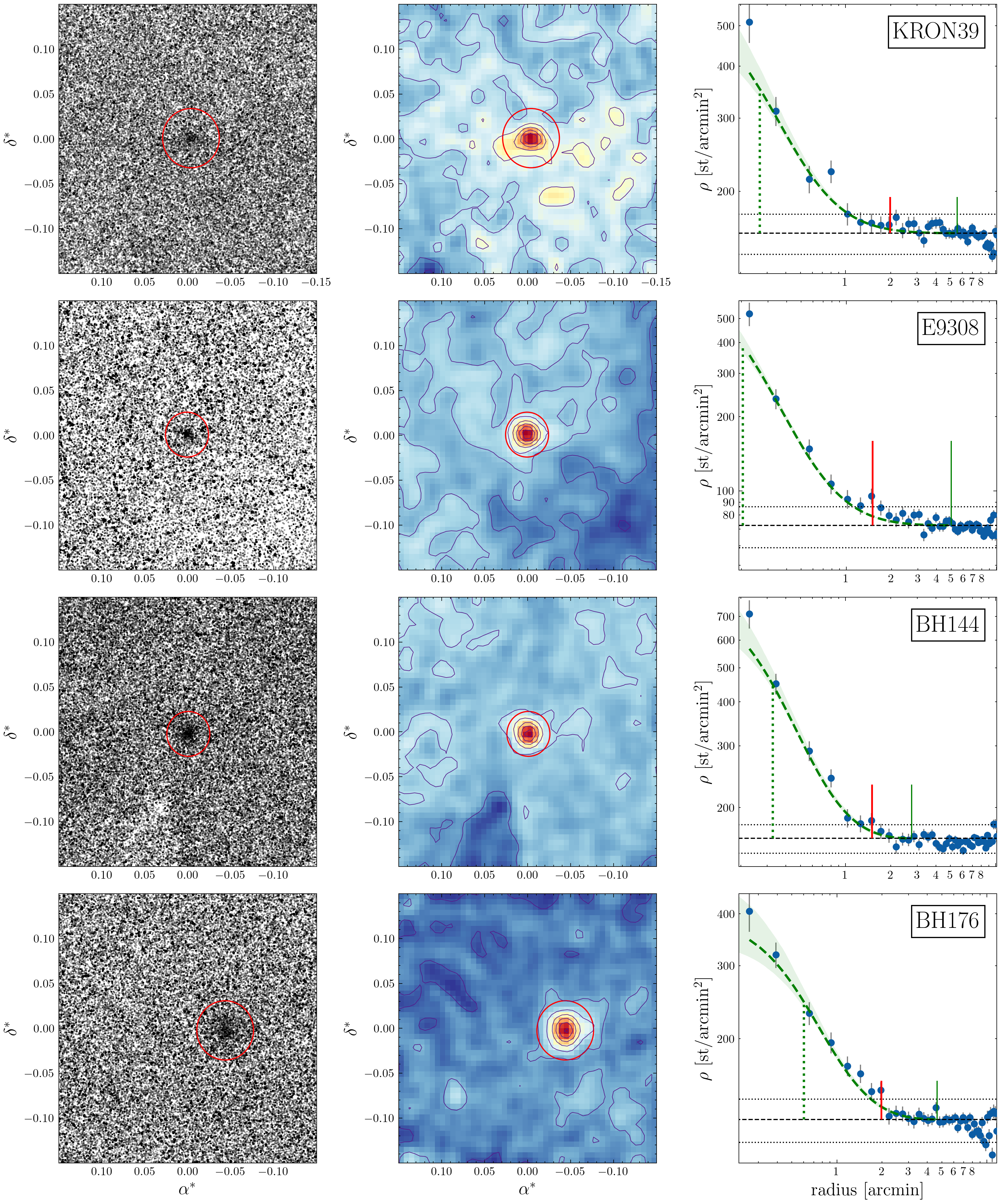}}
  \caption{Same as Fig.~\ref{fig:BER29_struct} for KRON39, E9308, BH144, and BH176.}
  \label{fig:16struct}
 \end{figure*}

\begin{figure*}
  \resizebox{\hsize}{!}{\includegraphics[]{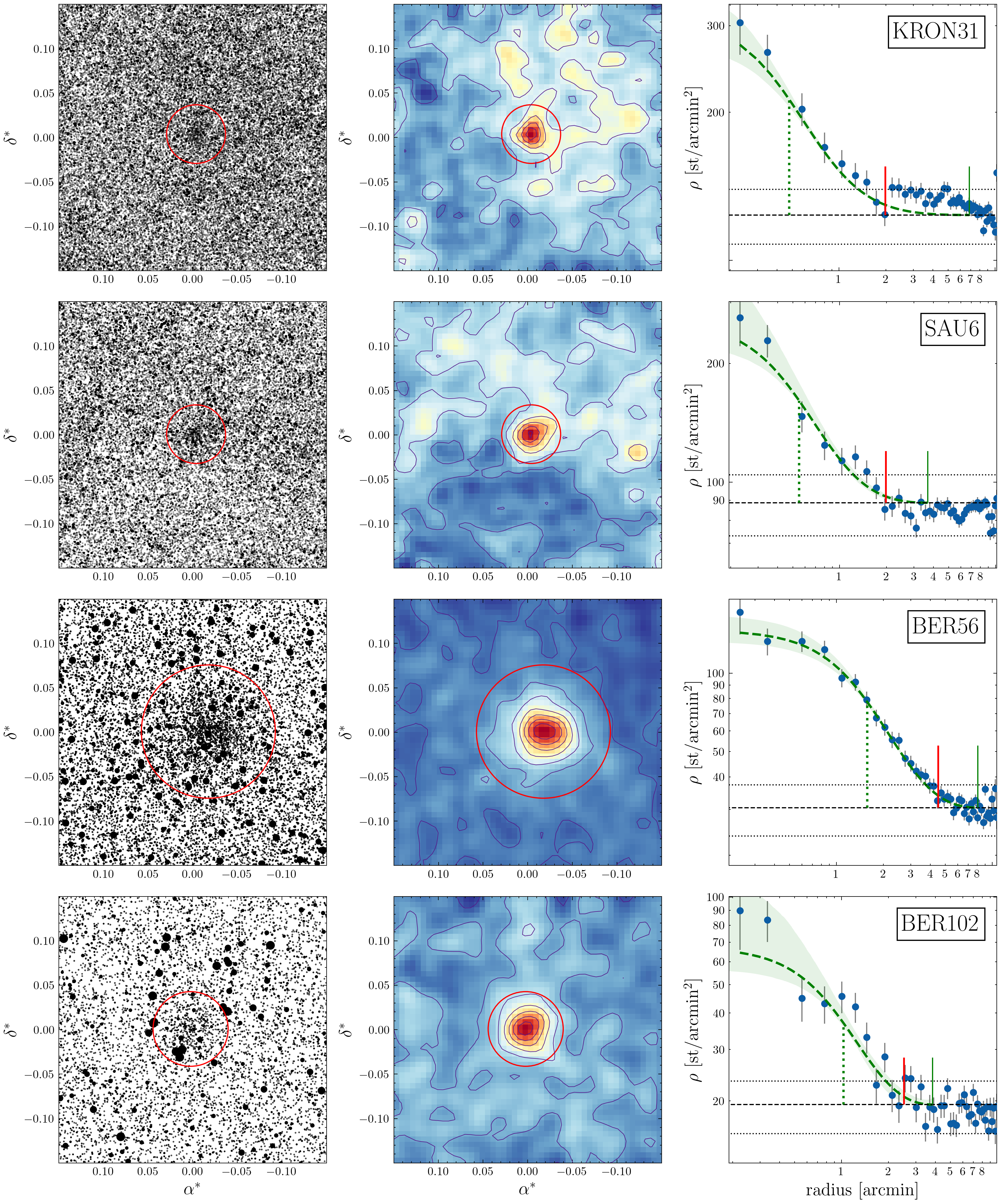}}
  \caption{Same as Fig.~\ref{fig:BER29_struct} for KRON31, SAU6, BER56, and BER102.}
  \label{fig:20struct}
 \end{figure*}

 We show the core, tidal and adopted radii used in the analysis in
 Table~\ref{tab:radii}. The figures showing the processed frame, density map,
 and density profile for all the clusters except Berkeley 29 are shown in
 Figs~\ref{fig:0struct}, \ref{fig:4struct}, \ref{fig:8struct}, 
 \ref{fig:12struct}, \ref{fig:16struct}, and \ref{fig:20struct}.

 \begin{table}[h!]
 \caption{Core ($r_{c}$), tidal ($r_{t}$), and adopted ($r_{a}$) radii values
 in arcminutes. The first two are shown with their respective 16th an 84th
 percentiles in parenthesis.}
 \label{tab:radii}
 \centering
 \begin{tabular}{llll}
 \hline\hline
 Cluster & $r_{c}$ (16th, 84th) &  $r_{t}$ (16th, 84th) & $r_{a}$\\
 \hline
  BER73         & 0.6 (0.5, 0.7) &  4.9 (3.8, 6.3) &  2.0\\
  BER25         & 1.5 (1.2, 1.8) &  7.0 (6.2, 8.0) &  5.0\\
  BER75         & 0.4 (0.3, 0.5) &  5.0 (3.6, 6.6) &  2.0\\
  BER26         & 0.7 (0.5, 1.0) &  3.4 (2.5, 4.5) &  1.6\\
  BER29         & 0.5 (0.4, 0.5) &  7.4 (6.4, 8.6) &  3.0\\
  TOMB2         & 0.9 (0.8, 0.9) &  6.7 (6.1, 7.4) &  3.5\\
  BER76         & 1.6 (1.2, 2.5) &  7.4 (5.7, 9.6) &  4.0\\
  F1212         & 0.8 (0.6, 1.1) &  8.8 (6.6, 10.7) & 3.0\\
  SAU1          & 0.7 (0.5, 1.0) &  3.9 (2.9, 5.4) &  2.0\\
  CZER30        & 0.6 (0.4, 0.7) &  6.4 (4.8, 8.2) &  2.5\\
  ARPM2         & 1.4 (1.0, 2.0) &  4.1 (3.4, 4.9) &  3.0\\
  BH4           & 0.4 (0.3, 0.5) &  5.7 (4.0, 7.1) &  2.0\\
  F1419         & 1.2 (0.9, 1.8) &  6.0 (4.3, 8.3) &  3.0\\
  BH37          & 1.2 (0.8, 1.9) &  3.8 (2.5, 5.6) &  2.0\\
  E9205         & 1.0 (0.9, 1.3) &  4.4 (3.7, 5.3) &  3.0\\
  E9218         & 0.6 (0.6, 0.7) &  6.5 (5.9, 7.3) &  3.0\\
  SAU3          & 0.5 (0.4, 0.6) &  4.9 (3.8, 6.4) &  2.0\\
  KRON39        & 0.3 (0.2, 0.3) &  5.6 (4.1, 7.0) &  2.0\\
  E9308         & 0.2 (0.2, 0.2) &  5.0 (4.2, 5.7) &  1.5\\
  BH144         & 0.3 (0.3, 0.4) &  2.8 (2.3, 3.4) &  1.5\\
  BH176         & 0.6 (0.5, 0.7) &  4.6 (3.8, 5.8) &  2.0\\
  KRON31        & 0.5 (0.4, 0.6) &  6.9 (5.7, 7.6) &  2.0\\
  SAU6          & 0.5 (0.4, 0.7) &  3.7 (2.9, 4.8) &  2.0\\
  BER56         & 1.6 (1.4, 1.7) &  8.1 (7.4, 8.9) &  4.5\\
  BER102        & 1.0 (0.8, 1.5) &  3.8 (3.0, 4.9) &  2.5\\
 \hline
 \end{tabular}
 \end{table}

\FloatBarrier
\section{Fundamental parameters}
 \label{app:fundam_params}

\begin{figure*}[t]
  % \resizebox{\hsize}{!}{\includegraphics[height=.9\textheight]{0_fpars.png}}
  \centering
  \includegraphics[height=.95\textheight]{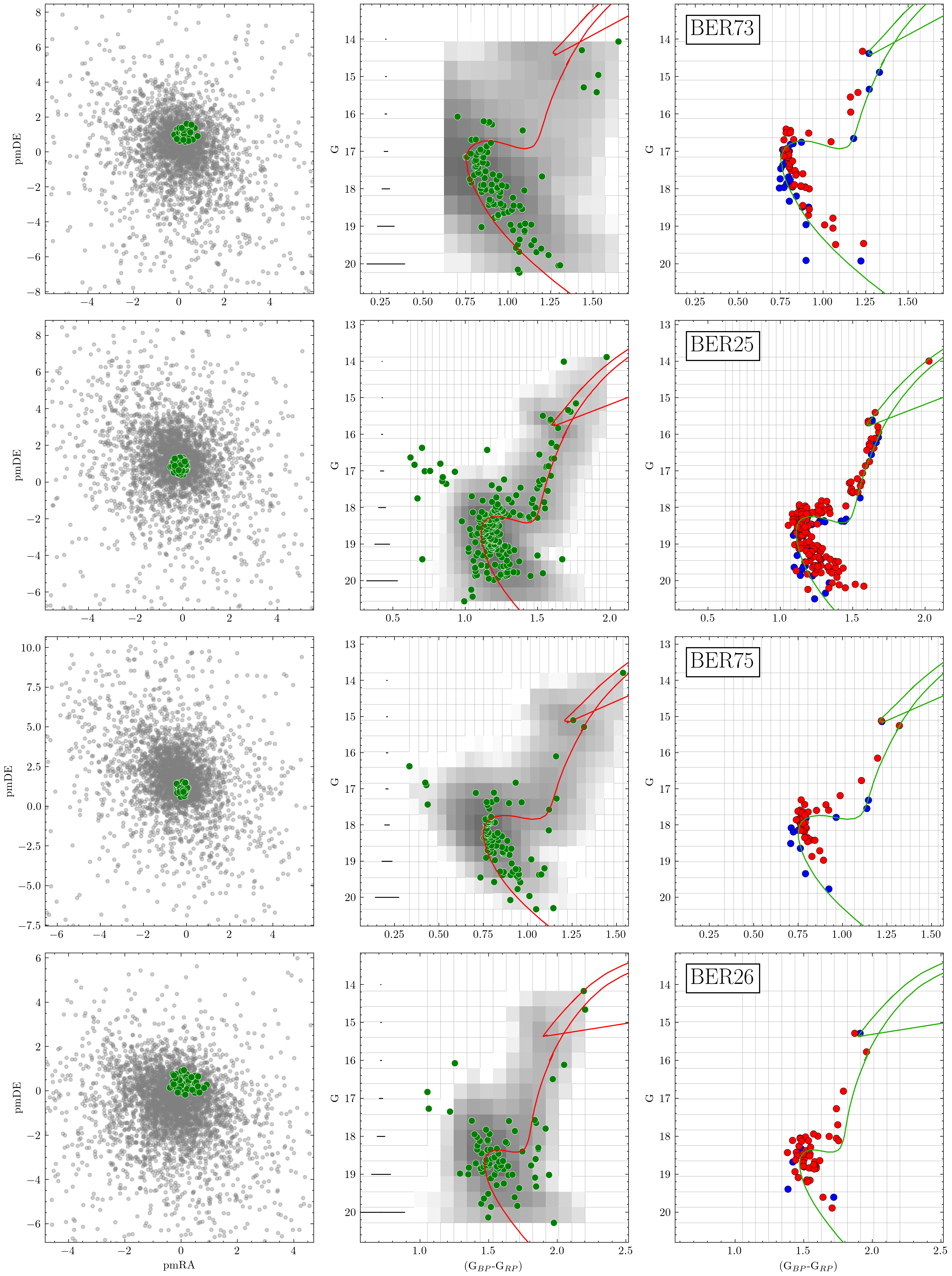}
  \caption{Same as Fig.~\ref{fig:BER29_fpars} for BER73, BER25, BER75, and BER26.}
  \label{fig:0fpars}
 \end{figure*}

\begin{figure*}
  % \resizebox{\hsize}{!}{\includegraphics[height=.9\textheight]{4_fpars.png}}
  \centering
  \includegraphics[height=.95\textheight]{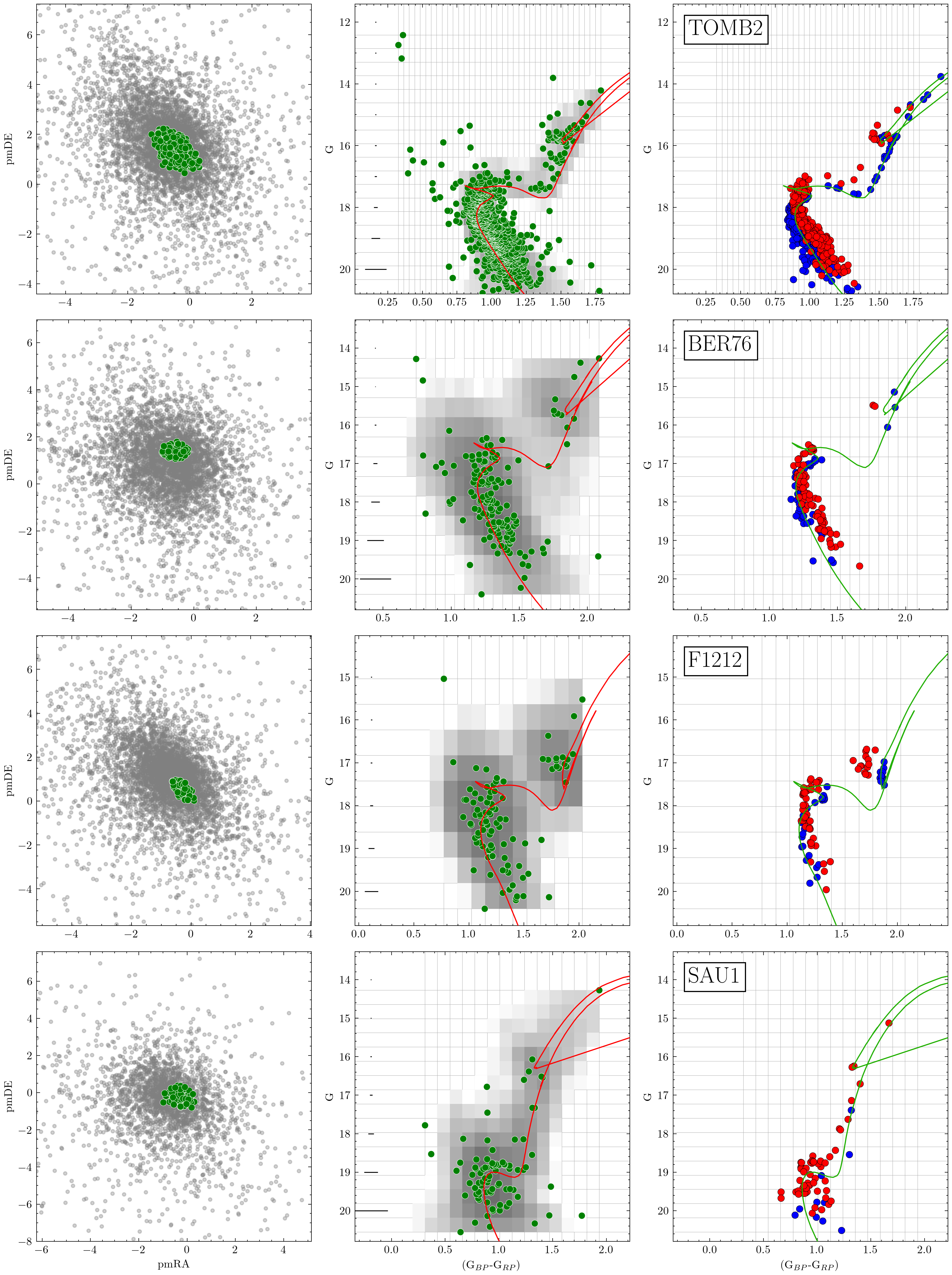}
  \caption{Same as Fig.~\ref{fig:BER29_fpars} for TOMB2, BER76, F1212, and SAU1.}
  \label{fig:4fpars}
 \end{figure*}

\begin{figure*}
  % \resizebox{\hsize}{!}{\includegraphics[height=.9\textheight]{8_fpars.png}}
  \centering
  \includegraphics[height=.95\textheight]{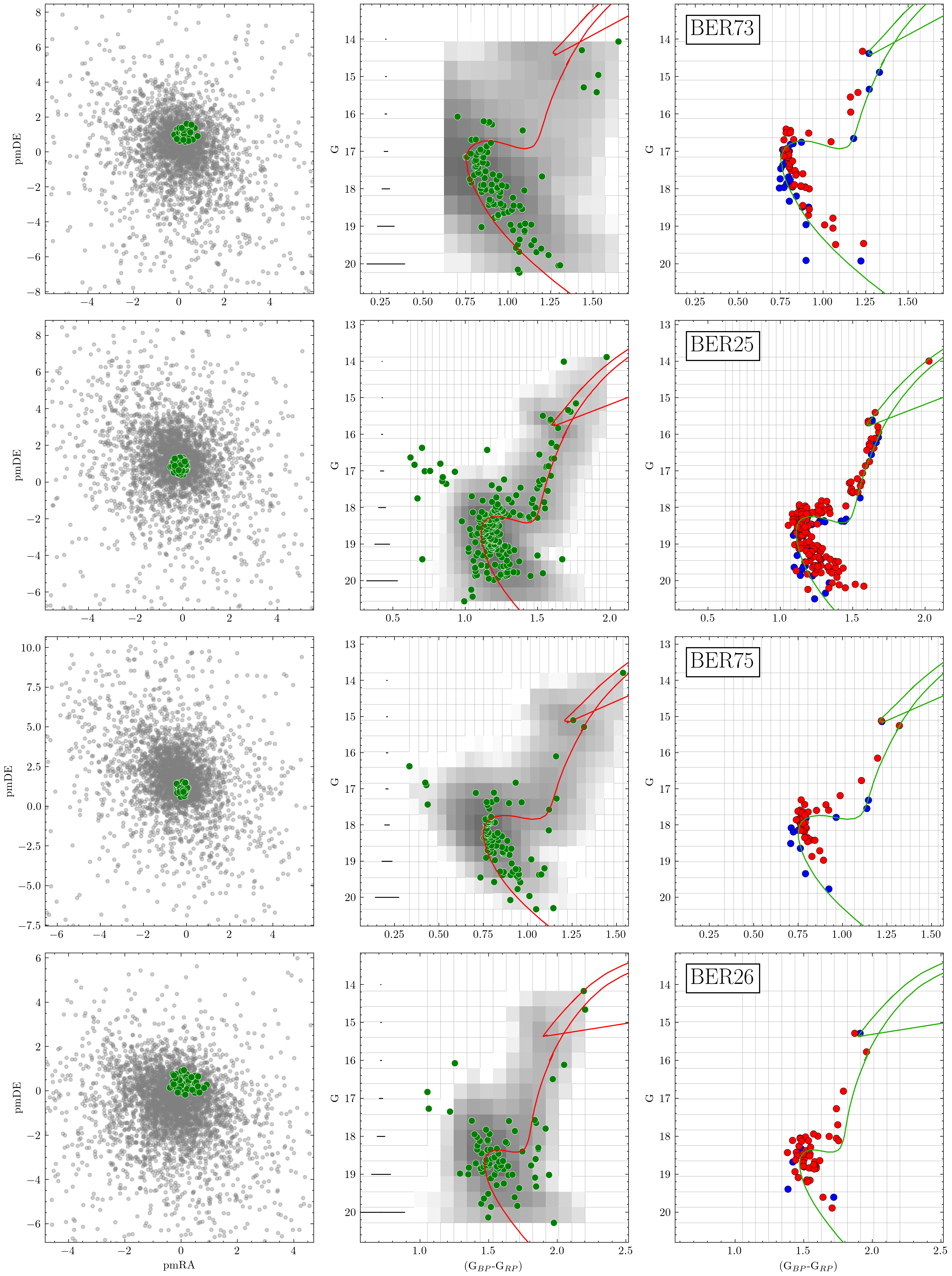}
  \caption{Same as Fig.~\ref{fig:BER29_fpars} for CZER30, ARPM2, BH4, and F1419.}
  \label{fig:8fpars}
 \end{figure*}

\begin{figure*}
  % \resizebox{\hsize}{!}{\includegraphics[height=.9\textheight]{12_fpars.png}}
  \centering
  \includegraphics[height=.95\textheight]{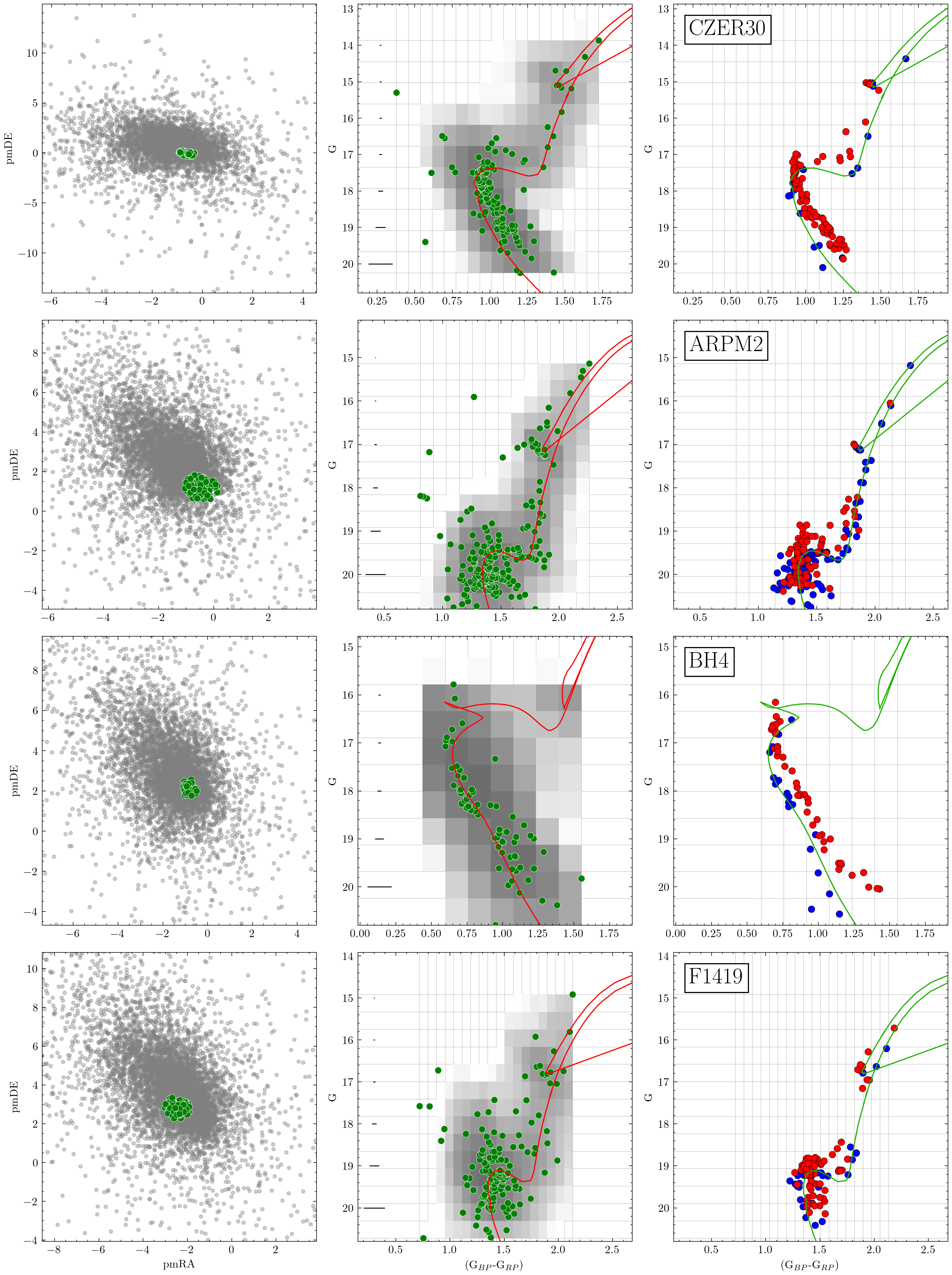}
  \caption{Same as Fig.~\ref{fig:BER29_fpars} for BH37, E9205, E9218, and SAU3.}
  \label{fig:12fpars}
 \end{figure*}

\begin{figure*}
  % \resizebox{\hsize}{!}{\includegraphics[height=.9\textheight]{16_fpars.png}}
  \centering
  \includegraphics[height=.95\textheight]{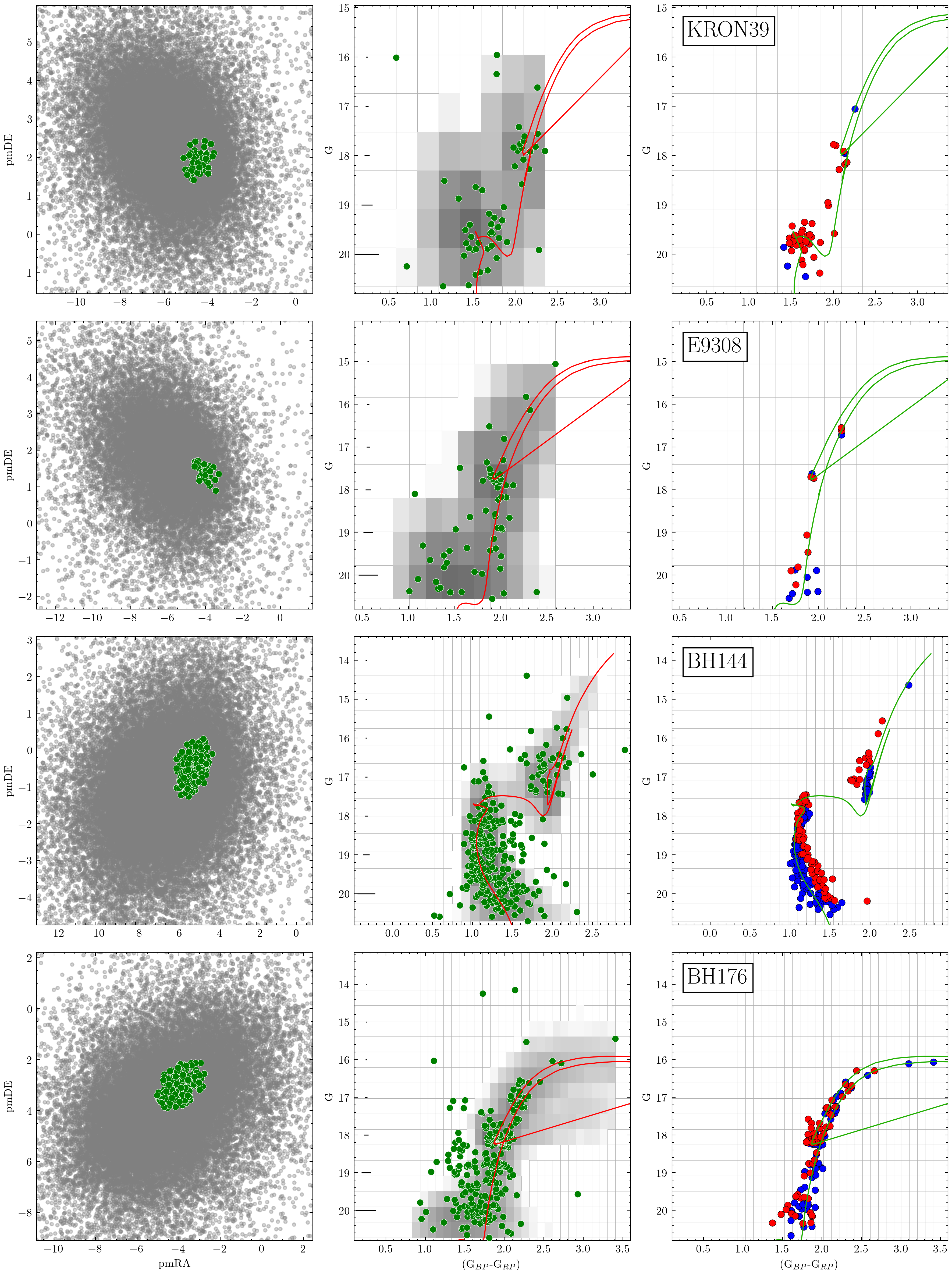}
  \caption{Same as Fig.~\ref{fig:BER29_fpars} for KRON39, E9308, BH144, and BH176.}
  \label{fig:16fpars}
 \end{figure*}

\begin{figure*}
  % \resizebox{\hsize}{!}{\includegraphics[height=.9\textheight]{20_fpars.png}}
  \centering
  \includegraphics[height=.95\textheight]{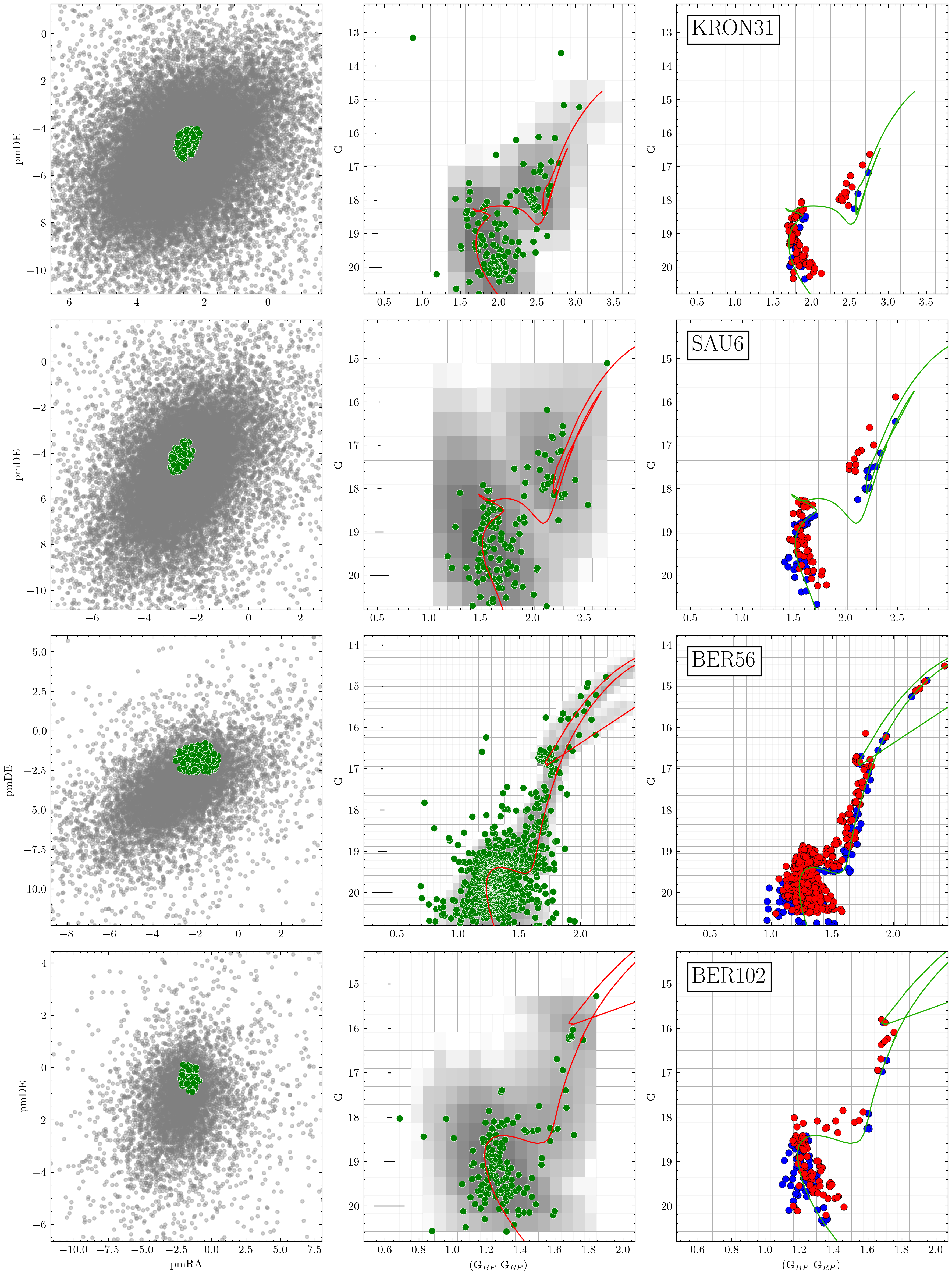}
  \caption{Same as Fig.~\ref{fig:BER29_fpars} for KRON31, SAU6, BER56, and BER102.}
  \label{fig:20fpars}
 \end{figure*}

 We show in Figs~\ref{fig:0fpars}, \ref{fig:4fpars}, \ref{fig:8fpars},
 \ref{fig:12fpars}, \ref{fig:16fpars}, and \ref{fig:20fpars} the vector-point
 diagram, CMD, and best match synthetic cluster for all the processed clusters
 except Berkeley 29.

\FloatBarrier
\section{Cluster by cluster discussion}
 \label{app:indiv_clusters}

  In this section we discuss each cluster separately, in the context of how our
  findings compare to those published in the literature.\\

  \noindent \textbf{Berkeley 73}: the CMD for this cluster shows a main sequence
  and a well defined turn-off (TO from now on) point, followed by a giant branch
  that is a bit redder than expected. Probably some stars above the TO could be
  candidates to blue stragglers. There is no strong evidence of stars forming a
  red clump (RC hereinafter).
  The results from~\cite{Ortolani_2005}, \cite{Carraro_2005},
  and~\cite{Carraro_2007_oldOC} for this cluster are 2.3/1.5/1.5 Gyr and
  6.5/9.7/11.5 kpc for the age and distance, respectively.
  \texttt{ASteCA} estimates a distance and age of 5.5 kpc and 4 Gyr,
  closer and older than the values found in any of the four databases and the
  above mentioned articles. Our distance is closer to the one reported
  in~\cite{Dias_2021} of 5.8 kpc, with an age of 2.2 Gyr. This is thus a
  rather old cluster, located well below the 9 Kpc limit.\\

  \noindent \textbf{Berkeley 25}: The CMD shows a clear giant branch and several stars
  above the TO that can be candidates to blue straggler stars. The RDP of
  Berkeley 25 shows a radius about $5^{\prime}$, the largest one in the sample.
  % These values are also rather different from the ones estimated
  % by~\cite{Carraro_2005}, who found an age of 3.0 Gyr and a distance of 11.3
  % kpc. The authors assumed a cluster radius of $0.8^{\prime}$, which is only
  % 20\% of the one we found.
  In~\cite{Carraro_2005} the authors assumed a cluster radius of $0.8^{\prime}$,
  only 20\% of the one we employed, and estimated an age of 3.0 Gyr and a
  distance of 11.3 kpc. These values are similar to the ones given
  in~\cite{Carraro_2007_oldOC} of 5 Gyr and 13.2 kpc for the age and the
  distance.
  The distance assigned by \texttt{ASteCA} is 7.4 kpc with an age of 5.2 Gyr,
  meaning that the cluster is located closer than the Carraro et al.
  values, and is found to be slightly older than what is shown in the
  databases.
  Along with a better parameter estimation performed by \texttt{ASteCA} using
  Gaia data, the rather small cluster region used may explain the differences
  between our parameters and those by Carraro et al. The databases MWSC,
  WEBDA, and OC02 all list distances larger than 11 kpc for this cluster. CG20
  on the other hand gives a value of 6.8 kpc, much closer to our estimate
  although associated to a considerably smaller age of 2.5 Gyr.\\

  \noindent \textbf{Berkeley 75}: This is a sparse cluster, with less than 100
  detected members. \cite{Carraro_2005} claim that this cluster possess
  a TO at $V= 17.5$ mag, a radius of $1^{\prime}$, an age of 3.5 Gyr and is
  located at a distance of 9.8 kpc.
  Our structural analysis yielded a $2^{\prime}$ radius with the TO set at
  $G=17.7$ mag. The giant branch is poorly populated though it shows a two-star
  RC. There are some stars above the sub-giant branch that could be explained
  by binary systems. \texttt{ASteCA} concludes that Berkeley 75 is 5.5 Gyr old
  cluster, placed at 8 kpc. This distance is similar to the one found by CG20
  of 8.3 Kpc, although the age assigned by CG20 is substantially smaller (1.7
  Gyr). The age assigned by OC02 is closer to ours (4 Gyr) but their distance is
  larger by $\sim$1 kpc.\\

  \noindent \textbf{Berkeley 26}: This cluster appears as a not so relevant overdensity
  projected against the background field. Although its RDP is well established
  and the radius is near $2^{\prime}$, the cluster TO is diffuse due to the
  scatter of stars and the likely presence blue straggler candidates. The
  giant branch is even less notorious.
  The position adopted for the TO is $G=18.5$ mag resulting in a distance of 4.6
  kpc and an age of 8.6 Gyr. \cite{Piatti_2010} claim that Berkeley 26 is 4 Gyr
  old and is placed at 4.3 kpc. While their distance value is close to ours, the
  age difference is important. This is related to the fact that \texttt{ASteCA}
  sets the cluster almost half a magnitude below the TO used by Piatti et al.,
  due to the presence of a large number of binary systems (which our code
  estimated to be around 70\%).
  WEBDA and MWSC locate the cluster at 4.3 kpc and 2.7 kpc, respectively, the
  latter assigning to it a very young age of $\sim$0.5 Gyr. The OC02 database
  includes this cluster with a distance of 12.5 kpc. This is apparently
  a mistake, since the original source for this value is \cite{Piatti_2010}.\\

  \noindent \textbf{Berkeley 29}: \cite{Tosi_2004} carried out a photometric analysis
  and found that this could be the most distant cluster in our galaxy to date.
  The parameters they attribute to this object are an age of 3.5 Gyr and a
  distance ranging from 11.2 to 14.4 kpc.
  Similar to our study, they compared the observed CMD of Berkeley 29 against a
  synthetic cluster set. The corresponding CMD produced by our method
  shows a visible TO between $18<G<19$ mag followed by a sub-giant and giant
  branches with a RC at $G=16.5$ mag or slightly less. We found that the
  distance to Berkeley 29 is 14.4 kpc and the age is 3.7 Gyr, in good
  coincidence with Tosi et al. The CG20 catalog indicates an age of 3 Gyr and a
  distance of 12.6 kpc, smaller that our estimate. The 4 Gyr and 13.4
  kpc values reported by \cite{Frinchaboy_2006} are also rather close to ours.
  As we will see this turns out to be the second most
  distant catalogued cluster so far, below vd Bergh-Hagen 176, if we measure
  the distance to the Sun. If we measure instead the galactocentric distance,
  then Berkeley 29 is indeed the most distant catalogued cluster as found
  by~\cite{Tosi_2004}, located at 22 kpc from the galactic center (see
  Table~\ref{tab:velocities}).\\

  \noindent \textbf{Tombaugh 2}: This is a populated cluster with almost 900 identified
  members. A very wide main sequence followed by a giant branch with a relevant
  RC are evident in its CMD. We speculate that part of the stars
  above the TO point may be blue stragglers. Tombaugh 2 is 2.1 Gyr old and is
  placed at a distance of 8.7 kpc according to \texttt{ASteCA}. These values are
  close to the ones found in~\cite{Dias_2021} of 9 kpc and 2.3 Gyr, and
  to the ones given by CG20 which are 1.6 Gyr and 9.3 kpc for the age
  and distance, respectively. The distances reported by OC02 and MWSC are below
  $\sim$7 kpc, and the value found in WEBDA is above $\sim$ 13 kpc.
  \cite{Villanova_2010} claim that this cluster is at 7.2 kpc.
  Using a strategy of analysis supported by photometric arguments alone, 
  \cite{Frinchaboy_2008} proposed an abundance spread among the members of
  Tombaugh 2 (overlapping between poor and metal rich stars). They adopted a
  distance of 7.9 kpc and an age of 2.0 Gyr.
  % all values derived from previous photometric observations (see their article).
  Due to the presence of variable stars the cluster was also observed by
  \cite{Kubiak_1992}, estimating an age of 4 Gyr and a distance of 6.3 kpc.\\

  \noindent \textbf{Berkeley 76}: The CMD of Berkeley 76 shows a diffuse TO at $G=16.5$
  mag, followed by a relevant giant branch. The parameters found by
  \texttt{ASteCA} indicate a distance of 5.4 kpc, and an age of 1.8 Gyr, far
  from the 12.6 kpc distance given
  by \cite{Carraro_2013_Five} although the age they determined, 1.5 Gyr, is
  close to ours. This same distance is reported by OC02 and WEBDA, while a
  smaller and much more reasonable value of 4.7 kpc is given in CG20.
  The distance discrepancy can obey to the fact that Carraro et al. set
  the TO at $V=18.5$ mag, 2 magnitudes below ours.\\

  \noindent \textbf{FSR 1212}: This is a sparse cluster with less than 100 identified 
  members whose overdensity visibly stands out in the coordinates space. The CMD
  shows a defined cluster TO at $G=17.5$ mag with a giant branch and a RC that
  are also well established. We notice a slight reddening of stars along the
  giant branch that can be explained by the presence of binary systems 
  ($\sim$50\% for the cluster).
  Our analysis indicates an age around 1.3 Gyr and a distance of 10.1
  kpc. Both parameters are in good agreement with the CG20 database, which
  assigns an age of 1.4 Gyr and a distance of 9.6 kpc. MSWC on the other hand
  locates this cluster at 1.8 kpc with an age of 0.4 Gyr, which is entirely too
  close and too young.\\

  \noindent \textbf{Saurer 1}: First reported in~\citet[][along with Saurer 3 and Saurer 6]
  {Saurer_1994} this cluster was analyzed by \cite{Carraro_2003} under
  the name of Saurer A, who estimated an age near 5 Gyr and a distance of 13.8
  kpc for a TO placed at $V=19$ mag. According to these authors, Saurer 1 is the
  cluster with the largest galactocentric distance detected. In this work we
  found that this title actually belongs to Berkeley 29, which is $\sim2$
  kpc further out than Saurer 1 (see Table~\ref{tab:velocities}).
  Our CMD shows the position of the TO at $G=19$ mag, a visible red giant
  branch and a handful of stars assumed as RC stars. From our analysis we find
  that the age of Saurer 1 is 6.6 Gyr and its distance is 12.4 kpc. The
  disagreement with~\cite{Carraro_2003} is thus minimum regarding the
  cluster distance given the associated errors. A smaller age, 4.5 Gyr, and
  a barely larger distance, 13.1 kpc, were determined
  by~\cite{Frinchaboy_2006}.\\

  \noindent \textbf{Czernik 30}: The CMD for this cluster shows a robust main sequence
  and a defined giant branch. The TO is situated at $G=17.5$ mag approximately.
  Our analysis results in a distance of 6.5 kpc and an age of 3.6 Gyr. These
  values are in good agreement with those of~\cite{Dias_2021} (5.9 kpc, 3 Gyr),
  but the distance strongly differs with the one found in~\cite{Hayes_2015}.
  Hayes et al. determined 9.12 kpc and 2.8 Gyr for the distance and age,
  respectively, meaning that they have overestimated the cluster distance.
  Looking at their Table 7 we see that seven previous studies yielded distances
  in the range [7.9, 9.3] kpc, all of them larger than the new one derived by
  \texttt{ASteCA} in this work. This includes our own previous analysis of the
  cluster in~\cite{Perren_2015}, where we assigned a slightly larger distance of
  8 kpc.\footnote{Notice that the distance modulus value of 16.07 mag quoted in
  Table 7 of Hayes et al. is incorrect. The proper value is $\sim14.5$ mag, as
  shown in Table 7 of~\cite{Perren_2015}.}
  CG20, WEBDA, and MWSC all report similar distances to the one found here:
  6.6 kpc, 6.2 kpc, and 6.8 kpc, respectively. We conclude that this cluster
  is thus well below the 9.1 kpc distance listed in OC02.\\

  \noindent \textbf{Arp-Madore 2}: This cluster's CMD shows a well traced TO and
  red giant branch with a very short main sequence, and a
  well established RC at $G=17$ mag.
  % Looking at Fig. 7 of~\cite{Ortolani_1995} the TO is located
  % 17<V<18 and the TO at V=20 while ours is G=19.5.
  % The CMD we have produced shows a sort of blueshift suggesting the cluster is
  % less metallic than the sun as also stated Ortolani et al.
  %
  According to~\cite{Ortolani_1995} this cluster is located at a distance of
  12.4 kpc with an intermediate age (no specific value is given), and a metal
  content of [Fe/H]$\sim-0.3$.
  The UBVI CCD photometry analysis carried out in~\cite{Lee_1997} resulted in
  estimates of 8.87$\pm$0.65 kpc and $5\pm1$ Gyr for the distance and the age,
  and of [Fe/H]$=-0.51\pm0.12$ for the metallicity.
  CG20 estimated a distance of 11.7 kpc and an age of 3.0 Gyr.
  \texttt{ASteCA} found $\sim$11 kpc and 4.1 Gyr for the distance and
  age, respectively, with a metal content of [Fe/H]=-0.33 which matches the
  sub-solar values reported by Ortolani et al. and Lee.
  Our distance thus is a good match for that of CG20, falling within the range
  given by the two previous studies mentioned.\\

  \noindent \textbf{vd Bergh-Hagen 4}: With less than 70 confirmed members, this is the
  third less populated cluster. A very weak but extended main sequence is
  visible in the CMD stretching almost 4 magnitudes.
  % where stars above $G=18.5$ mag appear blue-shifted with respect to the main
  % sequence and those below this value placed along it.
  The cluster's distance and age are estimated to be 8.1 kpc and 1.3 Gyr,
  both with large associated uncertainties.
  \cite{Carraro_2007} performed VI CCD photometry on this object finding an
  age of 0.2 Gy and a distance of 19.3 kpc. Although the distance given by
  \texttt{ASteCA} after processing the Gaia data is not accurate (the 16th-84th
  range spans 3 kpc), the value given by Carraro et al. lies outside
  of the 2-sigma range making it highly unlikely. The age parameter on the other
  hand is much closer to our estimate.
  % Both distance and age are the most controversial points with these authors
  % since the Gaia data analysis leaves no chance that our distance can change for
  % over 10 kpc, and neither the age.
  Looking at Fig. 17 of Carraro et al. we can see that their CMD for this
  cluster is considerably more contaminated than ours, with a large number of
  obvious field stars polluting the diagram.\\

  \noindent \textbf{FSR 1419}: This is a poorly studied cluster first reported
  in~\cite{Froebrich_2007}. CG20 and MWSC estimated ages and distances of
  1.6/0.2 Gyr and 11.1/8.4 kpc, respectively.
  We see in the cluster's  CMD that the TO is located at $G\approx19$ mag,
  clearly immersed in a scattered region at the top or the cluster main sequence
  where some stars may be binaries while other are blue straggler candidates.
  Evidences of a RC are found by the star grouping at approximately G=16.5 mag.
  The age found in the present analysis is close to 4 Gyr and the distance is
  9.2 kpc. While the distance lies within the range defined by the MWSC and CG20
  databases, the age shows a large difference with both databases.\\  

  \noindent \textbf{vd Bergh-Hagen 37}: This cluster has been catalogued in OC02
  as an object placed a 11.22 kpc with an age about 0.7 Gyr. These values were
  revised in~\cite{Dias_2021} finding 3.4 kpc and 0.3 Gyr for the distance and
  age. There is a remarkable distance difference between both results. 
  Looking at our CMD we see a main sequence extending for over 4 magnitudes, no
  red giant branch nor evident RC stars. Our analysis gives 2.9 kpc distance and
  an age of about 0.7 Gyr, rather close to the values given
  in~\cite{Piatti_2010} who locate this object at a distance of 2.5 kpc with an
  age ranging from 0.7 to 1 Gyr. Taking into account that the OC02 data for this
  cluster comes from Piatti et al., there must have been a mistake when storing
  the value into the database.
  Large differences appear comparing our results with that from CG20, who
  report that vd Bergh-Hagen 37 is located at a distance of 4.0 kpc and its age
  is 0.17 Gyr.\\

  \noindent \textbf{ESO 092 05}: Using BVI photometry \cite{Ortolani_2008} estimated
  an age of 6 Gyr and a distance of 11 kpc for this object. Our distance is
  larger, 12.7 kpc, but the age is the same.
  % The TO from Ortolani et al is set near V=19 mag while ours is set  at few
  % more than G=18.5 mag.
  The CG20 analysis yielded a distance of 12.4 and an age of 6 Gyr, a good
  agreement with both Ortolani et al. and \texttt{ASteCA} results.
  The metallicity for this cluster is claimed to be markedly sub-solar in
  Ortolani et al., estimating it at $\sim$-0.7. We obtain a
  larger metal content of [Fe/H]$\sim$-0.12. Although these values are rather far
  apart, we note that in Fig. 5 of Ortolani et al. the metallicity can be
  estimated to be very close to our value if the extinction is set to
  $\sim$0.11 instead of the 0.17 value given by Ortolani et al. The former is
  the value estimated by \texttt{ASteCA} for $E_{BV}$, while the latter
  is almost the maximum extinction value for the region given
  by~\cite{Schlafly_2011}. The difference in metal abundance is probably a
  result of an overestimation of the cluster's extinction in Ortolani et al.\\

  \noindent \textbf{ESO 092 18}: Early work by~\cite{Kubiak_1991} employing BVI CCD
  photometry estimated $\sim$10 kpc and 8 Gyr for the distance and age of this
  cluster.
  % The VI observations by~\cite{Phelps_1994_develop} obtained 0.6 kpc and about 1
  % Gyr age.
  According to photometric analysis made by~\cite{Carraro1995}, this is a 5 Gyr
  cluster located at 8.1 kpc from the Sun.
  \texttt{ASteCA} was able to estimate its parameters from a robust 2 mag long
  main sequence where the TO is at G=18.5 mag, and the red giant branch and the
  RC are well defined. Some stars in the range 15$<G<$17.5 mag are candidates to
  being blue stragglers.
  % It is probable that this cluster has a large number of
  % binaries to explain the large number of star above the TO.
  Our analysis indicates an age of 4.8 Gyr, a reasonable match with the 2.9 Gyr
  value given by CG20 and slightly older than the values given in the rest of
  the databases. The estimated distance by \texttt{ASteCA} is of 11.2 Kpc, which
  coincides within the uncertainties with CG20 (9.9 kpc), OC02 (10.6 kpc), and
  MWSC (9.5 kpc), but is more than 10 kpc beyond the 600 pc listed in WEBDA.
  This value is most likely an error in WEBDA where the correct reference is
  not \cite{Phelps_1994_develop} but~\cite{Janes_1994}. In the latter article
  the distance given to this cluster is 6.3 kpc which is still far from our
  estimate, but much more reasonable than the 600 pc listed.\\

  \noindent \textbf{Saurer 3}: This is a poorly studied cluster also known as Saurer C.
  Its CMD shows a dispersed sequence, suggesting a TO point at about G=19 mag
  and an evident RC at G=16.5 mag. Members appear scattered around the short
  main sequence, probably because of increasing photometric errors of Gaia data
  for G$>$19 mag.
  The analysis with \texttt{ASteCA} gives a distance 6.1 kpc, an age of 6.5
  Gyr, and the largest binary fraction of all the analyzed clusters 
  ($\sim$86\%).
  \cite{Carraro_2003} utilized CCD VI photometry to derive a distance of 9.5
  kpc, over 3 kpc above our estimate, and an age of $\sim$2 Gyr, also rather
  different from our value. These differences in age and distance likely
  arise from wrong membership assignations when using photometric
  arguments. Although their CMD displays a larger sequence than
  ours, spanning around 2 magnitudes up to V=22 mag (their Fig 16), it can
  clearly be seen to suffer from severe field star contamination.
  The distance stored in the MSWC database is 7.1 kpc, the closest catalogued
  value to our own estimate. This is another cluster that is located below the 9
  kpc limit, in contradiction to the $\sim$9.5 kpc distance reported in
  OC02 (whose source is the Carraro \& Baume article).\\

  \noindent \textbf{Kronberger 39}: This cluster is immersed in a region of high field
  star contamination. It was studied in the JHK bands by~\cite{Kronberger_2006}
  and catalogued as a \emph{``cluster candidate with RC''}. Using a radius of 0.8
  arcmin (ours is 2 arcmins), these authors computed a distance of about 11.1
  kpc (the distance value used in the OC02 database) with no age estimate.
  Recently, it was selected to be studied in~\cite{Monteiro_2020} but
  discarded because their analysis either did not reveal an identifiable cluster
  sequences or the isochrone fit was poor.
  With only 55 selected members this is the less populated cluster in our
  sample. We identified the cluster TO near G=20 mag and the RC at
  G$\approx$17.5 mag and obtained a distance of 13 kpc, an acceptable
  agreement withe the Kronberger et al. estimate, and an age 2.8 Gyr.
  The values for the age and distance in the MWSC database are 1 Myr and 4.4
  kpc, very much in disagreement with the estimates given by \texttt{ASteCA}.\\

  \noindent \textbf{ESO 093 08}: With only 60 identified members this is the second less
  populated cluster in our sample. Using VI photometry~\cite{Bica_1999} located
  this cluster at a distance of 13.7 kpc, with an age in the range 4-5 Gyr.
  More recently, also using VI CCD photometry, \cite{Phelps_2003} estimated 14
  kpc and 5.5 Gyr for these two parameters, along with a sub-solar metal
  abundance in the range $-0.60\leq$[Fe/H]$\leq-0.20$.
  Given the limiting magnitude of the G observations in the Gaia EDR3 survey, we
  were only able to analyze the RC and a poorly populated red giant branch.
  % It is interesting that when seen at the CMD (their Fig 3) shown by
  % Bica et al.  the cluster RC is notorious and placed at 18 < V < 19. Ours is found
  % along the giant branch at almost G=18.
  \texttt{ASteCA} found a distance of 13.3 kpc, similar to the estimates from
  Bica et al. and Phelps \& Schick as well as the values in the OC02 (14 kpc)
  and MWSC (13.8 kpc) databases.
  The age of 8.1 Gyr assigned to this cluster in our analysis is very
  large, making ESO 093 08 the oldest cluster in our sample. The metal
  abundance estimated by \texttt{ASteCA} is also sub-solar ([Fe/H]$\sim$-0.32)
  in agreement with the Phelps \& Schick range. It is worth noting that WEBDA
  lists a small distance of 3.7 kpc for this cluster whose source is not
  clear.\\

  \noindent \textbf{vd Bergh-Hagen 144}: This cluster, also referred to as Andrew-Lindsay
  1 and ESO 96-SC04, is known for hosting the planetary nebula PHR
  1315-6555~\citep{Parker_2011}.
  In one of the first studies of this cluster~\cite{Janes_1994} estimated a
  distance of 7.6 kpc, very close to the value in the MWSC database. On the
  other end of the distance spectrum,
  \cite{Carraro_2005_neglected} assumed a cluster radius of $0.6^{\prime}$ and
  found a very large distance of 16.9 kpc, and an age of 0.8 Gyr. Looking at
  their Fig. 9 (lower panel) it is easy to see that determining cluster members
  is a complex task. Our analysis of the RDP suggests that the cluster's radius
  is over 1.5$^{\prime}$. The CMD shows a densely populated main sequence ending
  in a TO situated at G=17.5 mag. The giant branch is scattered but suggests a
  RC at the same magnitude value.
  % The presence of a large number of binaries cloud be the reason of
  % such a data scatter.
  Processing the selected cluster members with \texttt{ASteCA} results in a
  distance estimate of 10.1 kpc, rather different from the values
  from both Janes \& Phelps and Carraro et al., and an age around 1 Gyr.
  CG20 found a distance of 9.6 kpc and an age close to 1.4 Gyr,
  while~\cite{Majaess_2014} reported a distance of 10.0$\pm$0.4 kpc and an
  age of 0.8$\pm$3 Gyr. Both age and distance values for these two articles are
  very close to ours. Recently~\cite{Fragkou_2019} suggested a slightly larger
  distance of $\sim12$ kpc, the value present in OC02 and WEBDA, and an age of
  about 0.66 Gyr.\\

  \noindent \textbf{vd Bergh-Hagen 176}: Originally reported as an open cluster
  in~\cite{vandenBergh_1975}, this is a very interesting object that has been
  profusely studied over the past almost five decades. Much of the interest
  comes from the fact that it is still not settled whether this is a metal-rich
  globular cluster, an old open cluster, or a transition-type cluster in between
  these two objects.
  According to~\cite{Ortolani_1995} this is an object in the border line
  between globular and galactic clusters, located at a distance of 13.4 kpc
  and as old as NGC 6791 (i.e., in the range 4-8 Gyr).
  \cite{Phelps_2003} claim that this is a either a massive old cluster or a
  young globular cluster, in both cases metal-rich with [Fe/H]$=0.0\pm0.2$.
  The distance and age given in this work are 18$\pm$1 kpc and 7.0$\pm$1.5 Gyr,
  respectively.
  \cite{Frinchaboy_2006} analyzed this cluster in relation to the Galactic
  anticenter stellar structure (GASS), and obtained a distance of 15.8$\pm$0.5
  kpc, an age of 6.3$\pm$1 Gyr, and suggested a solar metal abundance in
  coincidence with Phelps \& Schick.
  \cite{Davoust_2011} used 2MASS and FORS2 VLT photometry and labeled this an
  old metal-rich open or transition-type cluster. The estimated  distance is
  15.1$\pm$0.5 kpc with an age in the range 6-7 Gyr, and a metal content of
  [Fe/H]$\sim-0.10\pm0.1$.
  In~\cite{vandenBergh_2011} the author lists this object as part of the known
  Galactic globular clusters. The values are taken from
  the~\cite{Harris_1996,Harris_2010} catalog of globular
  clusters,\footnote{\url{https://physics.mcmaster.ca/~harris/mwgc.dat}}
  where the distance is given as 18.9 kpc,
  and the metallicity is solar ([Fe/H]=0.0) with no age value reported.
  Using medium-resolution Gemini spectroscopy~\cite{Sharina_2014} obtained
  values of 15.2$\pm$0.2 kpc, 7$\pm$0.5 Gyr, and [Fe/H]=-0.1$\pm$0.1, for the
  distance, age, and metallicity, respectively. The authors conclude that this
  is an old metal-rich open cluster that could belong to the thick disk.
  Recently \cite{Vasiliev_2021} used Gaia EDR3 parallaxes of 90 selected
  members of this cluster to derive a distance of 13.9$\pm$3.5 kpc
  (0.072$\pm$0.018 mas). They state that this cluster is part of a group of
  objects that may be old open clusters rather than globular clusters.
  Although the position of the TO can not be established as it is not visible
  in the CMD, the RC for this cluster is clearly defined which allowed
  \texttt{ASteCA} to estimate a distance of 18.3 kpc and a cluster age of 5
  Gyr. Our distance is thus far from the Vasiliev et al. value, but close to the
  value reported in the Harris catalog. It is also very similar to the value
  given by the MWSC catalog of $\sim19$ kpc (with an age of 6.3 Gyr).
  This large heliocentric distance means that, if confirmed as an open
  cluster, vd Bergh-Hagen 176 could be the most remote open cluster found to
  date. In agreement with previous studies \texttt{ASteCA} also classifies this
  cluster as metal rich ([Fe/H]$\approx$0.15) and massive (M$\approx170000\,
  M_{\odot}$), the most massive object of our sample by far.\\

  \noindent \textbf{Kronberger 31}: This poorly studied and compact cluster was reported
  in~\cite{Kronberger_2006} and classified as a \emph{``cluster candidate with
  RC''}. Using a $1.3^{\prime}$ radius to delimit the cluster region, the authors
  assigned an excess of $E_{BV}$=0.84 mag and a distance of 11.9 Kpc.
  It is possible to identify in the CMD a short main sequence, the TO at G=18.3
  mag and the giant branch. The data scatter around the giant branch can be
  explained by the effect of field star contamination, large color excess in the
  region (\texttt{ASteCA} found $E_{BV}\approx1.3$ mag, $\sim$0.5 mag larger
  than the Kronberger et al. value), and the presence of a significant number of binary
  systems (almost 80\% according to \texttt{ASteCA}). Our analysis resulted in
  distance and age values of 7.6 kpc and 1 Gyr, respectively, which
  locates this cluster below the 9 kpc limit in disagreement with Kronberger et
  al. and the MWSC catalog that assigns a distance of 12.6 kpc.\\

  \noindent \textbf{Saurer 6}: This is another poorly studied cluster whose CMD denotes a
  wide and short main sequence, with a TO showing at approximately G=18 mag. The
  giant branch is quite scattered although the RC is evident.
  The RDP gives a cluster radius of $2^{\prime}$ with a high star density
  present in the region, which may explain the main sequence widening.
  This object was studied by~\cite{Frinchaboy_2002} using CCD VI photometry,
  resulting in estimates of 2 Gyr and 9.3 kpc for the age and distance.
  Both values are not far from \texttt{ASteCA}'s results for these parameters
  which are 1.4 Gyr and 9.2 kpc. WEBDA and OC02 report the same distance, but
  the MWSC catalog assigns a smaller distance of 7.3 kpc.\\

  \noindent \textbf{Berkeley 56}: Always classified as an old open
  cluster~\citep[see for example][]{King_1964}, this is a well populated object
  and the second cluster with the largest number of identified members in our
  sample. It is also the second most massive cluster in our sample according
  to \texttt{ASteCA} with $\sim53000\,M_{\odot}$.
  \cite{Janes_1994} estimated its heliocentric distance at 5.7 kpc and an age of
  5.67 Gyr~\citep[according to][]{Salaris_2004}. \cite{Carraro_2006} found a
  much larger distance of 12.1 kpc and an age of 4 Gyr. \cite{Janes_2011} on the
  other hand found an even larger distance of 15.2 kpc and a slightly larger age
  of 6 Gyr.
  Both OC02 and WEBDA list the Carraro et al. values for both parameters, while
  MWSC and CG20 report 13.2 kpc and 2.5 Gyr, and 9.5 kpc and 3
  Gyr, respectively. There is undoubtedly a substantial spread for both
  parameters across recent studies for this cluster.
  \texttt{ASteCA} estimates 11.1 kpc and 5.2 Gyr for the distance and age,
  both values within the range defined by the above mentioned articles.\\

  \noindent \textbf{Berkeley 102}: This is a curious cluster regarding its distance
  estimates in the literature. The four catalogs used in this work list wildly
  different values: 2.6 kpc~\citep[WEBDA;][]{Tadross_2008}, 4.9 kpc (MWSC),
  9.6 kpc~\citep[OC02;][]{Hasegawa_2008}, 10.5 kpc (CG20). Likewise, the
  assigned ages in these catalogs range from 0.6 Gyr to $\sim4$ Gyr.
  \cite{Maciejewski_2008} analyzed BV and 2MASS photometry and concluded that
  this is \emph{``only a chance alignment of physically unrelated star''}.
  The CMD for Berkeley 102 shows a scattered main sequence with the TO at G=18.5
  mag and a well defined giant branch, which leads us to reject the conclusion
  by Maciejewski \& Niedzielski and classify this object as a true open cluster.
  \texttt{ASteCA} places Berkeley 102 at 7.4 kpc and assigns an age of 4.9 Gyr.
  The age estimate is close to that of CG20 (3.9 Gyr) but the distance is
  almost 3 kpc smaller, locating this cluster below the 9 kpc limit.\\
  % The
  % difference between our distance estimate and that of CG20 probably arises from
  % the large binary fraction found by \texttt{ASteCA} of almost 60\%.\\
  % There is a small
  % color excess discrepancy between our analysis and that of CG20 that cannot
  % reduce the gap between results. Certainly, when looking at the CMD, one is led
  % to think that we could set the TO at G=18 instead of G=18.5, but this would
  % end up in an even shorter distance and therefore larger discrepancy with CG20.

\FloatBarrier
\section{Galactocentric and dynamical parameters}
  \label{app:galac_dynam}

  Galactocentric values used to construct~Fig~\ref{fig:MWmap_vectors} are shown
  in Table~\ref{tab:velocities}.

  \begin{table*}
  \caption{Galactocentric coordinates and velocities (proper motions, radial)
  for each analyzed cluster. References for the radial velocities: [1]: 
  \cite{Tarricq_2021}, [2]: \cite{Dias_2002}, [3]: \cite{Soubiran_2018}, [4]: 
  \cite{Dias_2007}, [5]: \cite{Frinchaboy_2006}. The values marked with an
  asterisk ($^*$) where obtained for E9218 and KRON31 from a single member for
  each cluster (from our own selection) from Gaia EDR3 data.}
  \label{tab:velocities}
  \centering
  \renewcommand{\arraystretch}{1.3}
  \begin{tabular}{lcccccccc}
   \hline \hline
   $Cluster$ & $R_{CG}$ & $X$ & $Y$ & $Z$ & $\mu_{\alpha}$ & $\mu_{\delta}$ & $RV$\\
    & [kpc] & [kpc] & [kpc] & [kpc] & [mas/yr] & [mas/yr] & [km/s]\\
   \hline
   BER73 & $12.96$ & $-12.55$ & $-3.13$ & $-0.87$ & $ 0.23 \pm 0.17$ & $ 1.06 \pm 0.20$ & $112.41$ [1] \\
   BER25 & $14.19$ & $-13.11$ & $-5.28$ & $-1.21$ & $- 0.13 \pm 0.14$ & $ 0.87 \pm 0.18$ & $108.07$ [1] \\
   BER75 & $14.32$ & $-12.72$ & $-6.40$ & $-1.53$ & $- 0.22 \pm 0.12$ & $ 1.14 \pm 0.17$ & $122.41$ [1] \\
   BER26 & $12.35$ & $-12.16$ & $-2.12$ & $0.22$ & $ 0.16 \pm 0.28$ & $ 0.38 \pm 0.26$ & $68.00$ [2]\\
   BER29 & $22.23$ & $-21.69$ & $-4.40$ & $2.06$ & $ 0.15 \pm 0.28$ & $- 1.05 \pm 0.27$ & $25.72$ [1]\\
   TOMB2 & $15.07$ & $-13.36$ & $-6.90$ & $-1.01$ & $- 0.49 \pm 0.23$ & $ 1.39 \pm 0.28$ & $122.47$ [1]\\
   BER76 & $12.53$ & $-11.93$ & $-3.83$ & $-0.16$ & $- 0.60 \pm 0.18$ & $ 1.41 \pm 0.14$ & $73.02$ [1]\\
   F1212 & $16.65$ & $-14.92$ & $-7.38$ & $-0.46$ & $- 0.34 \pm 0.15$ & $ 0.52 \pm 0.20$ & $71.82$ [1]\\
   SAU1 & $19.56$ & $-18.20$ & $-6.98$ & $1.64$ & $- 0.29 \pm 0.29$ & -$ 0.25 \pm 0.27$ & $98.00$ [2]\\
   CZER30 & $13.48$ & $-12.62$ & $-4.72$ & $0.51$ & -$ 0.62 \pm 0.12$ & $ 0.07 \pm 0.11$ & $82.07$ [1]\\
   ARMP2 & $15.83$ & $-12.17$ & $-10.07$ & $-1.09$ & $- 0.49 \pm 0.23$ & $ 1.25 \pm 0.26$ & $58.25$ [1]\\
   BH4 & $13.29$ & $-10.88$ & $-7.57$ & $-0.98$ & $- 0.82 \pm 0.14$ & $ 2.12 \pm 0.16$ & -- \\
   F1419 & $12.89$ & $-9.07$ & $-9.09$ & $-1.04$ & $- 2.48 \pm 0.24$ & $ 2.85 \pm 0.21$ & -- \\
   BH37 & $8.96$ & $-8.50$ & $-2.83$ & $-0.07$ & $- 3.50 \pm 0.09$ & $ 3.99 \pm 0.11$ & $51.90$ [3]\\
   E9205 & $13.05$ & $-4.62$ & $-12.09$ & $-1.65$ & $- 3.00 \pm 0.29$ & $ 2.48 \pm 0.23$ & $57.40$ [1]\\
   E9218 & $11.78$ & $-4.84$ & $-10.67$ & $-1.29$ & $- 3.59 \pm 0.36$ & $ 2.72 \pm 0.27$ & $65.47$* \\
   SAU3 & $8.81$ & $-6.53$ & $-5.90$ & $0.34$ & $- 6.70 \pm 0.24$ & $ 3.29 \pm 0.19$ & -- \\
   KRON39 & $12.85$ & $-3.77$ & $-12.27$ & $-0.44$ & $- 4.42 \pm 0.35$ & $ 1.89 \pm 0.23$ & -- \\
   E9308 & $12.48$ & $-2.85$ & $-12.12$ & $-0.93$ & $- 4.04 \pm 0.24$ & $ 1.38 \pm 0.17$ & $86.00$ [2]\\
   BH144 & $8.53$ & $-2.31$ & $-8.19$ & $-0.55$ & $- 5.14 \pm 0.33$ & -$ 0.47 \pm 0.32$ & $40.00$ [4]\\
   BH176 & $12.14$ & $7.40$ & $-9.53$ & $1.35$ & $- 3.97 \pm 0.45$ & -$ 3.07 \pm 0.36$ & $11.20$ [5]\\
   KRON31 & $8.06$ & $-4.53$ & $6.66$ & $0.26$ & $- 2.36 \pm 0.15$ & -$ 4.62 \pm 0.29$ & $32.10$* \\
   SAU6 & $9.72$ & $-4.69$ & $8.51$ & $0.47$ & $- 2.60 \pm 0.18$ & -$ 4.15 \pm 0.29$ & -- \\
   BER56 & $13.31$ & $-7.35$ & $11.05$ & $-0.99$ & $- 1.92 \pm 0.32$ & $- 1.81 \pm 0.32$ & $-54.95$ [1] \\
   BER102 & $12.90$ & $-10.99$ & $6.74$ & $-0.59$ & $- 1.56 \pm 0.25$ & $- 0.36 \pm 0.21$ & -- \\ 
   \hline
   \end{tabular}
  \end{table*}

\end{appendix}

\end{document}